\documentclass{article}

\usepackage{arxiv}

\usepackage[utf8]{inputenc} 
\usepackage[T1]{fontenc}    
\usepackage{hyperref}       
\usepackage{url}            
\usepackage{booktabs}       
\usepackage{bm}
\usepackage{amsfonts}       
\usepackage{amsmath}
\usepackage{nicefrac}       
\usepackage{microtype}      
\usepackage{lipsum}
\usepackage{graphicx}
\graphicspath{ {./images/} }
\usepackage{comment}
\usepackage{amsmath}
\usepackage{natbib}
\usepackage{rotating}
\usepackage{multirow}
\usepackage{hyperref}

\DeclareMathOperator*{\argmax}{arg\,max}
\newcommand*\colvec[3][]{
    \begin{pmatrix}\ifx\relax#1\relax\else#1\\\fi#2\\#3\end{pmatrix}
}

\title{Optimal risk-assessment scheduling for primary prevention of cardiovascular disease}

\author{
 Francesca Gasperoni \\
  MRC Biostatistics Unit\\
  University of Cambridge\\
  Cambridge, U.K. \\
  \texttt{francesca.gasperoni@mrc-bsu.cam.ac.uk} \\
   \And
Christopher H. Jackson \\
  MRC Biostatistics Unit\\
  University of Cambridge\\
  Cambridge, U.K. \\
  \texttt{chris.jackson@mrc-bsu.cam.ac.uk} \\
  \And
 Angela M. Wood \\
  Department of Public Health and Primary Care\\
  University of Cambridge\\
  Cambridge, U.K.\\
  \texttt{amw79@medschl.cam.ac.uk} \\
  \And
 Michael J. Sweeting \\
  Department of Health Sciences\\
  University of Leicester\\
  Leicester, U.K.\\
  \texttt{michael.sweeting@leicester.ac.uk} \\
  \And
 Paul J. Newcombe \\
  MRC Biostatistics Unit\\
  University of Cambridge\\
  Cambridge, U.K. \\
  \texttt{paul.newcombe@mrc-bsu.cam.ac.uk} \\
  \And
 David Stevens \\
  Liverpool Centre for Cardiovascular Science\\
  University of Liverpool\\
  Liverpool Heart \& Chest Hospital\\
  Department of Cardiovascular and Metabolic Medicine\\
  Institute of Life Course and Medical Sciences\\
  University of Liverpool\\
  Liverpool, U.K.\\
  \texttt{david.stevens@liverpool.ac.uk} \\
  \And
 Jessica K. Barrett \\
  MRC Biostatistics Unit\\
  University of Cambridge\\
  Cambridge, U.K. \\
  \texttt{jessica.barrett@mrc-bsu.cam.ac.uk} \\
}

\begin{document}
\maketitle
\begin{abstract}
In this work, we introduce a personalised and age-specific Net Benefit function, composed of benefits and costs, to recommend optimal timing of risk assessments for cardiovascular disease prevention.  We extend the 2-stage landmarking model to estimate  patient-specific CVD risk profiles, adjusting for time-varying covariates. We apply our model to data from the Clinical Practice Research Datalink, comprising primary care electronic health records from the UK. We find that people at lower risk could be recommended an optimal risk-assessment interval of 5 years or more. Time-varying risk-factors are required to discriminate between more frequent schedules for higher-risk people.
\end{abstract}


\section{Introduction}
\label{s:intro}
The World Health Organization identified cardiovascular disease (CVD) as the leading cause of morbidity and mortality across the world, with 17.9 million deaths from CVD in 2016 (31\% of all global deaths, \citet{whoreport}).
The prescription of statins and other lipid-lowering medication is recognised as the most common primary prevention strategy for CVD~\citep{reiner2013statins} with the UK National Institute for Health and Care Excellence (\citet{niceguidelines}) guidelines recommending offering atorvastatin $20$ mg to people who have a $10\%$ or greater 10‑year risk of developing CVD. 
The 10-year CVD risk is recommended to be computed through the QRISK2 assessment tool~\citep{hippisley2008predicting} every 5 years from age 40 for both men and women. However, there is no universal agreement on the best risk-assessment strategy~\citep{pylypchuk2018cardiovascular,lalor2012guidelines,arnett20192019,piepoli20162016}.
In particular, identifying the optimal CVD risk-assessment frequency is an open problem, as recognised  by~\citet{piepoli20162016}: 
"[repeating CVD risk-assessment occasionally], such as every 5 years, is recommended, but there are no data to guide this interval".  

The problem of optimal timing for risk-assessment is crucial in preventive medicine and it is widely studied in cancer screening~\citep{bibbins2016screening,shieh2017breast,ito2019screening}, but is much less investigated in CVD risk-assessment~\citep{selvarajah2013identification, lindbohm20195}. The optimal risk-assessment schedule is often identified via the maximization of a Utility function~\citep{rizopoulos2016personalized, sweeting2017using} or via the minimization of a Cost function~\citep{bebu2018optimal}. A third option is represented by the Net Benefit function defined as the difference between benefits and costs~\citep{gray2011applied}. These functions are tailored to the specific disease of interest, as they are composed of quantities that are considered discriminatory for that particular condition. 
Elements evaluated for building these functions might include: the quality-adjusted life years (QALYs) gained, the expected life years gained, the cost associated with a risk-assessment, the expected number of risk-assessments and the undetected time spent with an undiagnosed condition.
Furthermore, the optimal risk-assessment schedule could depend on the \textit{stage} of the disease of interest, as for~\citet{bebu2018optimal}, or on the specific \textit{risk level} of developing the disease, as for~\citet{lindbohm20195}. To deal with the dynamic nature of the problem, multi-state models~\citep{bebu2018optimal,lindbohm20195} and joint models~\citep{rizopoulos2016personalized, sweeting2017using} have been investigated. 
But only a few authors have provided personalised recommendations for the next screening~\citep{rizopoulos2016personalized, sweeting2017using, bebu2018optimal}.

In this work we introduce a \textit{personalised} and \textit{age-specific} monitoring schedule that aims to provide an optimal balance between benefits and costs associated with statins initiation. Our recommendations are based on evidence obtained from large-scale Electronic Health Records (EHR) data. Considering the size of the data and its complexity (i.e., sparse repeated measurements and missing values), joint models and multi-state models would be computationally unfeasible. Instead, we exploit the landmarking framework described by~\citet{van2011dynamic} and at each specific landmark age, we maximize a person-specific Net Benefit (NB) function.
The elements that characterise the Net Benefit functions are: the CVD Free Life Years gained over a 10 year time horizon (as benefit); the expected number of visits, the cost associated with a CVD event, the cost of statin consumption (as costs). The idea of considering CVD Free Life Years and cost of statins for defining a Net Benefit function was proposed by \citet{rapsomaniki2012framework} in a different context (they proposed the NB as an alternative measure for comparing different risk prediction models).  
A key element in the proposed NB function is the statin initiation time for each person, at each landmark. In order to estimate the statin initiation, we have to define a dynamic CVD risk profile for each person at each landmark. The risk profile is estimated by extending the two stage landmarking approach by \citet{paige2018landmark}. Specifically, we extend the first stage, by providing not only Best Linear Unbiased current predictions through a linear mixed effect model with random intercept and slope, but also Best Linear Unbiased future predictions of time-varying risk factors for CVD onset. Exploiting future predictions enables better informed risk-assessment strategies compared to those based only on current risk factors.

The paper is organised as follows. The motivating dataset is described in Section~\ref{s:data}. The proposed model and method is presented in Section~\ref{s:mm}. The results obtained for men and women separately are shown in Section~\ref{s:results}. 
The final discussion is reported in Section~\ref{s:discussion}.

\section{Motivating data}
\label{s:data}
Our motivating dataset is derived from the Clinical Practice Research Datalink (CPRD), which covers approximately 6.9\% of the UK population and is representative in terms of age and gender~\citep{herrett2015data}. This dataset is linked to secondary care admissions from Hospital Episode Statistics (HES), and national mortality records from the Office for National Statistics (ONS) \citep{herrett2015data}. 
The linked dataset is composed of $2,610,264$ patients and $39,189,729$ measurements (i.e., Body Mass Index, high lipoprotein cholesterol, systolic blood pressure, smoking status, total cholesterol).\\
We exclude those people with prevalent CVD or statin treatment before study entry. We also exclude individuals who had no measurements of any of BMI (Body Mass Index), SBP (Systolic Blood Pressure), total cholesterol, HDL (High Density Lipoprotein) cholesterol, or smoking status between study entry and study exit dates. \\
We include as risk factors the following continuous variables: Body Mass Index (BMI), systolic blood pressure (SBP), total cholesterol, HDL cholesterol; and the following binary variables: smoking status (current smoker or not), statin consumption index, blood pressure medication index, diagnoses of diabetes, renal disease, depression, migraine, severe mental illness, rheumatoid arthritis and atrial fibrillation. We also include the Townsend deprivation index as a categorical variable with 20 levels. These risk factors are chosen because they are part of the QRISK2~\citep{hippisley2008predicting} and QRISK3~\citep{hippisley2017development} risk scores. \\ 
CVD is defined as any of the following: acute myocardial infarction, stroke, angina or 
transient ischemic attack, in line with the definition used in the QRISK3 CVD risk score \citep{hippisley2017development}.\\ 
A total of 1,971,002 individuals (914,951 men and 1,056,051 women) from 406 GP practices were included in the study. We randomly allocated 2/3 of practices (270 practices with 1,774,220 individuals) to the derivation cohort dataset and 1/3 of practices (136 practices with 836,044 individuals) to the validation cohort dataset.
Further details about risk factors and outcome definitions and cohort selection are reported in section 1 of the supplementary material.

\section{Models and methods}
\label{s:mm}
We introduce and optimize a Net Benefit (NB) function in which we account for both benefit and costs associated with risk-assessments and statins initiation.
The NB function is defined as a \textit{age} and \textit{person-specific} function, whose optimization leads to the identification of a personalised risk-assessment schedule for primary prevention of CVD, at each \textit{age} of interest. We define the ages of interest as our \textit{landmark ages}, $L_a =\{ 40, 45, 50, …,80\}$. This choice mimics the current visit schedule recommended by~\citet{niceguidelines}. At each landmark age, we select those people in the derivation set who have not been diagnosed with CVD, are still alive and have not yet received statins, defined as the \textit{landmark cohort}.
Each landmark age represents the \textit{time origin} for the NB evaluation, while $L_a + 10$ represents the \textit{time horizon} (years is the scale), i.e.,  we consider a potential CVD risk-assessment frequency from every year to every 10 years. 

A key point in the model definition is statins initiation, assumed to happen at the first risk-assessment scheduled after their 5-year CVD risk exceeds the 5\% threshold. Following~\citet{van2013efficiency}, we evaluate 5-year CVD risk instead of 10-year CVD risk due to lack of follow-up. 
The expected time of crossing the 5\% threshold is landmark \textit{age} and \textit{person-specific}, and it is denoted as $t_{i,L_a}^*$.
Statins initiation has a positive impact on lengthening the CVD-free life years~\citep{ferket2012personalized} and it has been proved to reduce the risk of a CVD event by about 20\% as reported by a previous meta-analysis of statin trials~\citep{unit2005efficacy}. Then, we model this effect via the hazard ratio, $\theta$, that is set equal to $0.8$. Further details on the definition of CVD-free life years are given in Section~\ref{ss:LMEM}.

All analyses are run separately on men and women in the derivation set, since incidence of CVD is substantially higher in men than women.

\subsection{Net benefit}
\label{ss:nb}
For each person $i$ at $L_a$, we define the optimal risk-assessment strategy, $\boldsymbol{\tau}^{opt}_{i,L_a}$, as that which maximises the Net Benefit function among a set of $F$ risk-assessment strategies of interest ($\boldsymbol{\tau}^f \in \{\boldsymbol{\tau}^1, \boldsymbol{\tau}^2, .., \boldsymbol{\tau}^F\}$) as in Eq.~\eqref{eq:max_INB}.
\begin{equation}
    \boldsymbol{\tau}_{i,L_a}^{opt} = \argmax_{f \in \{1,..,F\}}{NB_{i,L_a}(\boldsymbol{\tau}^f)}.
    \label{eq:max_INB}
\end{equation}
The risk-assessment schedule, $\boldsymbol{\tau}^f$, is a vector of visit times $\boldsymbol{\tau}^f = \{\tau_1^f,\tau_2^f,..,\tau_V^f\}$, characterised by $f$, the time between two visits (i.e., $f = 1$ stands for yearly evaluation). Note that the visit times are all fixed in advance and are defined by common time intervals and $\tau_1^f$ is always equal to the origin time, $L_a$, while $\tau_V^f \leq L_a+10$.
The risk-assessment scheduled after person $i$ is expected to have a 5-year CVD risk higher than 5\%, $t_{i,L_a}^*$ at landmark age $L_a$, is denoted as $\tau_{k_{i,L_a}^*}^f$.  To avoid overly heavy notation, we drop the superscript $f$ in the remaining part of this section. 


$NB_{i,L_a}(\boldsymbol{\tau})$ is defined in monetary terms, by converting health outcomes to the scale of costs, and subtracting the actual costs of health service usage, if required.  Health outcomes are measured as expected quality-adjusted CVD-free life years (QALYs), over a maximum time of 10 years, and can be converted to expected costs by multiplying the amount $\lambda$ that a decision-maker is willing to pay for one year of full health. We assume that $\lambda$ ranges from £20,000/year to £30,000/year~\citep{niceguidelines}.
The expected costs are composed of all costs associated with the expected CVD-free life years of a person (up to a maximum of 10 years), including the yearly cost of statins taken after $\tau_{k_{i,L_a}^*}$ and the expected costs of risk-assessment visits.

Firstly, the $NB_{i,L_a}(\boldsymbol{\tau})$ is defined in Eq.~\eqref{eq:nb_screening}.  
\begin{equation}
NB_{i,L_a}(\boldsymbol{\tau}) = QALY(\boldsymbol{\tau}) \cdot \lambda - cost(\boldsymbol{\tau}).
\label{eq:nb_screening}    
\end{equation}
$QALY(\boldsymbol{\tau})$ is based on  the following elements: 
\begin{itemize}
    \item $EFLY_{NS}(\tau_{k_{i,L_a}^*})$: Time \textit{before statin initiation} spent free of CVD, or event-free life years, EFLY without statins. This time can be computed as the integral of the probability of not developing CVD, with no statins initiation, between time origin and $\tau_{k_{i,L_a}^*}$.
    \item $EFLY_{S}(\tau_{k_{i,L_a}^*})$: Time \textit{after statin initiation} spent free of CVD, or event-free life years, EFLY with statins. This time can be computed as the integral of the probability of not developing CVD, after statins initiation, between $\tau_{k_{i,L_a}^*}$ and time horizon.
\end{itemize}
We assume that $EFLY_{NS}(\tau_{k_{i,L_a}^*})$ is associated with a utility equal to 1 (full health), while $EFLY_{S}(\tau_{k_{i,L_a}^*})$ is associated with a disutility $u_s$. Statins are considered to be very low risk drugs, associated with a utility reduction from 0 to 0.003, given to \textit{pill burden} \citep{kong2018decision}. 
This means that $u_{s} \in [0.997,1]$. 
Refer to Eq.~\eqref{eq:qualy_strategy} for the extended definition of $QALY(\boldsymbol{\tau})$.

\begin{equation}
    QALY(\boldsymbol{\tau}) = EFLY_{NS}(\tau_{k_{i,L_a}^*}) + u_s \cdot EFLY_S(\tau_{k_{i,L_a}^*}).
    \label{eq:qualy_strategy}
\end{equation}

The expected costs associated with a predefined risk-assessment strategy $\boldsymbol{\tau}$ are composed of the yearly cost of statins, $c_s$ [\pounds/year], taken for $EFLY_S(\tau_{k_{i,L_a}^*})$ years, the expected costs of visits, defined as the cost of a single visit $c_{\nu}$ [\pounds/visit] multiplied by the expected number of visits, $\mathbb{E}_{\boldsymbol{\tau}}[N_i]$: 
\begin{equation}
  cost(\boldsymbol{\tau}) = c_s\cdot EFLY_S(\tau_{k_{i,L_a}^*}) + c_{\nu} \cdot \mathbb{E}_{\boldsymbol{\tau}}[N_i].
  \label{eq:cost_strategy}
\end{equation}
The cost of statins per year of life, $c_s$ ranges from £4.3/year to £321.2 /year, assuming  a daily dose of 20 mg of Atorvastatin~\citep{nicefarma}. The cost of a single visit, $c_{\nu}$ is assumed to be £18.39/visit \citep{kypridemos2018future}. To estimate $\mathbb{E}_{\boldsymbol{\tau}}[N_i]$, we assume that the CVD risk-assessments are performed up to time $\tau_{k^*_i}$ (i.e., no more visits after statins initiation). An example of the $\mathbb{E}_{\boldsymbol{\tau}}[N_i]$ estimate is reported in seection 2.2 of the supplementary material.


By combining the Equations between~\eqref{eq:nb_screening} and~\eqref{eq:cost_strategy}, we are able to define the Net Benefit function for the $i$-th person at landmark age $L_a$, associated with a specific risk-assessment schedule $\boldsymbol{\tau}$.

To compute Eq.~\eqref{eq:nb_screening}, we estimate the $EFLY_{NS}(\tau_{k_{i,L_a}^*})$,  $EFLY_S(\tau_{k_{i,L_a}^*})$, $EFLY_{NS}$ and $\tau_{k_{i,L_a}^*}$. The Event Free Life Years are estimated through a 2-stage landmarking approach where the event of interest is a CVD diagnosis between $L_a$ and $L_a + 10$ (see Section~\ref{ss:LMEM}). Considering the definition of $\tau_{k^*_{i,L_a}}$ as the first visit after  $t^*_{i,L_a}$, the problem collapses to the prediction of $t^*_{i,L_a}$.  We provide an extended 2-stage landmarking model in Section~\ref{ss:cross} to estimate $t^*_{i,L_a}$. 
We set the values of $\lambda$, $u_s$, $c_s$, $c_{\nu}$ in Section~\ref{s:results} and run a sensitivity analysis to assess the robustness of our analysis with respect to the variability of these parameters (section 5 in the supplementary material).

Two special cases of Eq.~\eqref{eq:max_INB} can be identified. The first one is when the 5-year CVD risk of a person is not expected to cross the 5\% threshold in the time-window of interest $[L_a, L_a +10]$, which means $\tau_{k^*_{i,L_a}} \geq L_a + 10$. In this case, Eq.~\eqref{eq:max_INB} is driven term related to the expected number of visits. The optimal risk-assessment strategy is therefore the one associated with the lowest expected number of visits $\mathbb{E}_{\boldsymbol{\tau}}[N_i]$. The second case is when the 5-year CVD risk is predicted higher than the 5\% threshold at $L_a$, which means $\tau_{k^*_{i,L_a}} = L_a$. We are not interested in evaluating this case, as these people should initiate statins already at $L_a$.

\subsection{Two-stage landmarking approach for CVD risk prediction}
\label{ss:LMEM}
In this Section, we describe how we apply the two-stage landmarking model proposed by~\citet{paige2018landmark} in order to estimate the probability of not being diagnosed with CVD, before statins initiation. 

At each landmark age $L_a \in \{40,45,..,80\}$, we fit a Linear Mixed Effect Model (LMEM) with random intercepts and slopes to all individuals from the derivation dataset who are in the \textit{landmark cohort}. The outcomes of interest are the time-varying risk factors for CVD.  
Let $smoke_{ij}$, $SBP_{ij}$, $TCHOL_{ij}$, $HDL_{ij}$, $BMI_{ij}$, $BPM_{ij}$, $statin_{ij}$ and $age_{ij}$ denote the repeated measures of smoking status, systolic blood pressure, total cholesterol, HDL cholesterol, body mass index,  history of blood pressure-lowering medication, statin prescription and age for individual $i$, $i \in \{1,.., N_{L_a}\}$, recorded at visit $j$, $j \in  \{1,.., J_i\}$, where $N_{L_a}$ is the landmark cohort size.
In order to get the most precise estimates of the model regression parameters, we include not only past measurements taken prior to the landmark age, but also future measurements~\citep{paige2018landmark}. The LMEM is defined as:

\begin{equation}
\label{eq:lmem}
\begin{aligned}[b]
    smoke_{ij} &= \beta_{10}  + \beta_{11} age_{ij} + u_{10i} + u_{11i}age_{ij} + \varepsilon_{ij}  \\ 
    HDL_{ij}     &= \beta_{20}  + \beta_{21} age_{ij} + u_{20i} +u_{21i}age_{ij} + \varepsilon_{ij}   \\
    SBP_{ij}     &= \beta_{30}  + \beta_{31} age_{ij} + \beta_{32} BPM_{ij} + u_{30i} + u_{31i}age_{ij} + \varepsilon_{ij}\\
    TCHOL_{ij}   &= \beta_{40}  + \beta_{41} age_{ij} +  \beta_{42} statin_{ij} + u_{40i} + u_{41i}age_{ij} + \varepsilon_{ij}  \\
    BMI_{ij} &= \beta_{50}  + \beta_{51} age_{ij} + u_{50i} + u_{51i}age_{ij} + \varepsilon_{ij} 
\end{aligned}
\end{equation}
where $\colvec{\textbf{u}_{0i}}{\textbf{u}_{1i}} \sim MVN(\textbf{0}, \Sigma)$ and $\Sigma$ is a full matrix; $\boldsymbol{\varepsilon}_{ij} \sim MVN(\textbf{0}, \boldsymbol{\sigma}_e I)$ and $I$ is the identity matrix. 
Here $\boldsymbol{\beta}_0$ represents fixed intercepts for each risk factor, $\boldsymbol{\beta}_1$ represents fixed slope for each risk factor. $\beta_{32}$ represents an adjustment factor in systolic blood pressure levels for those subjects under blood-pressure lowering medication at the time the measurement was taken. $\beta_{42}$ is the regression parameter that represents the effect of statin prescription on total cholesterol.
$\textbf{u}_{0i}$ and $\textbf{u}_{1i}$ are vectors of risk factor-specific random intercepts and random slopes respectively and are correlated between risk factors.
Finally, $\boldsymbol{\varepsilon}_{ij}$ represents uncorrelated residual errors for each risk factor.\\
Our model assumes that all risk factors jointly follow a multivariate normal distribution, which is plausible for BMI, SBP, total cholesterol, HDL cholesterol but less plausible for smoking status which is a dichotomous variable. However, inference based from the multivariate normal distribution may often be reasonable even if the multivariate normality does not hold~\citep{schafer1997analysis}.

We complete the first stage of the two-stage landmarking approach by predicting \textit{current risk factor values} using the Best Linear Unbiased Predictors (BLUPs) for each person $i$ of the landmark cohort, at time $L_a$ (denoted as $\widehat{SBP}_{iL_a}$, $\widehat{TCHOL_{iL_a}}$, $\widehat{smoke}_{iL_a}$, $\widehat{HDL}_{iL_a}$ and $\widehat{BMI}_{iL_a}$).
Importantly, we only take advantage of \textit{past observations} for computing the BLUPs because the prediction of the time-varying CVD risk factors should not depend on future information. The prediction of the BLUPs therefore mirrors the prediction as it would be carried out for a new individual who we have only observed up to the landmark age.

In the second stage of the landmarking approach, we fit a Cox proportional hazard model at each  landmark age. The event of interest is the time to CVD diagnosis over the next 10 years (people diagnosed with CVD after the time-horizon $L_a+10$ are censored).
The risk factors included in the Cox proportional hazard model at time $L_a$, are of two types: \textit{time-fixed} or \textit{time-varying}. The \textit{time-fixed} risk factors are diabetes, blood pressure medication, renal disease, depression, migraine, severe mental illness, rheumatoid arthritis, atrial fibrillation diagnosis and Townsend deprivation score. These risk factors are assumed known at the landmark age $L_a$ and are assumed to be constant over time from the landmark age. We denote these risk factor values for person $i$ as $\textbf{x}_{i,fixed}$. The \textit{time-varying} risk factors are  BMI, SBP, total cholesterol, HDL cholesterol and smoking status. The values included in the Cox model at time $L_a$ are the BLUPs \textit{resulting from the first stage}. We refer to these values for person $i$ as $\textbf{x}_{i,BLUP}(L_a)$.
We assume hazard at time $L_a$ in Eq.~\eqref{eq:coxmodel_paige}.
\begin{equation}
         \label{eq:coxmodel_paige}
\begin{aligned}[b]
        \lambda^{NS}(t;\textbf{x}_i(L_a),L_a)  
        &= \lambda^{NS}_0(t;L_a) \cdot \exp\Big\{\textbf{x}_{i,fixed}^T \boldsymbol{\beta}_{fixed}(L_a) +
        \textbf{x}_{i,BLUP}^T(L_a)   \boldsymbol{\beta}_{BLUP}(L_a)\Big\}. \\
\end{aligned}
\end{equation}
Given Eq.~\eqref{eq:coxmodel_paige}, we are able to estimate the probability a person will not be diagnosed with CVD by time t, given they
are not on statins, $S^{NS}(t;\textbf{x}_i(L_a), L_a)$, that is $\Lambda_0(t;L_a)\exp\{\textbf{x}_i(L_a)^T \boldsymbol{\beta}(L_a)\}$, where $\Lambda_0(t;L_a)$ is the cumulative baseline hazard and $\textbf{x}_i(L_a)$ is the vector of \textit{all} risk factors of a person $i$, measured at $L_a$. Following the definition given in the previous section, we can write the $EFLY_{NS}(\tau_{k_{i,L_a}^*})$ as in Eq.~\eqref{eq:efly_NS_extended}.
\begin{equation}
EFLY_{NS}(\tau_{k_{i,L_a}^*}) = \int_{L_a}^{\tau_{k^*_{i,L_a}}} S^{NS}(t;\textbf{x}_i(L_a), L_a) \, dt.
    \label{eq:efly_NS_extended}
\end{equation}
Analogously, the $EFLY_{S}(\tau_{k_{i,L_a}^*})$ is reported in Eq.~\eqref{eq:efly_S_extended}.
\begin{equation}
EFLY_{S}(\tau_{k_{i,L_a}^*}) = \int_{\tau_{k^*_{i,L_a}}}^{L_a+10} S^{S}(t;\textbf{x}_i(L_a), L_a) \, dt
    \label{eq:efly_S_extended}
\end{equation}
where $S^{S}(t;\textbf{x}_i(L_a), L_a)$ is the probability of not being diagnosed with CVD after statins initiation and is equal to $S^{NS}(\tau_{k^*_i};\textbf{x}_i(L_a)) \times \left( \frac{S^{NS}(t;\textbf{x}_i(L_a))}{ S^{NS}(\tau_{k^*_i};\textbf{x}_i(L_a))} \right)^{\theta}$. Complete details on the derivation of $S^{S}(t;\textbf{x}_i(L_a), L_a)$ can be found in section 2.1 of the  supplementary material.

\subsection{Extending the 2-stage model for predicting the 5\% crossing time}
\label{ss:cross}
We introduce the extension to the 2-stage landmarking model, required to provide the 5-year CVD personalised risk profile to predict $t^*_{i,L_a}$, conditional on the history of the person at landmark age $L_a$. 

Firstly, we define the \textit{prediction time set} at landmark age $L_a$ ($\mathcal{P}_{L_a} = \{L_a, L_a +1, L_a + 2,..,L_a +10\}$), as the collection of times at which we want to estimate the 5-year CVD risk after the current landmark age. 
The whole landmark cohort will not be alive after 1, 2, 3..,10 years and it would not be sensible to predict values for people that died or have been diagnosed with CVD before the time of interest $s$, $s \in \mathcal{P}_{L_a}$. Therefore, it is necessary to create a \textit{sub-cohort} composed only of those people that are still alive and are not diagnosed with CVD at each prediction time $s$, $s \in \mathcal{P}_{L_a}$.
Using the LMEM~\eqref{eq:lmem} fitted to individuals in the landmark cohort in Section~\ref{ss:LMEM}, we are able to compute $\widehat{SBP}_{is}$, $\widehat{TCHOL_{is}}$, $\widehat{smoke}_{is}$, $\widehat{HDL}_{is}$ and $\widehat{BMI}_{is}$ as the Best Linear Unbiased Predictors (BLUPs) for each person $i$ belonging to the \textit{landmark sub-cohort}, at each time $s$ in the prediction time set $\mathcal{P}_{L_a}$.

Given the BLUPs computed at each time $s \in \mathcal{P}_{L_a}$, we fit a Cox proportional hazard model at each time $s$, on the landmark sub-cohort. We are interested in 5-year CVD risk prediction, so all events happening later than $s + 5$, $s \in \mathcal{P}_{L_a}$, are considered as censored at time $s+5$.
We use a Cox proportional hazard model fitted at each prediction time $s$, $s \in \mathcal{P}_{L_a}$ in Eq.~\eqref{eq:coxmodel}. This equation is identical to Eq.~\eqref{eq:coxmodel_paige}, apart from (i) the origin time $s$ ($L_a$ in the previous section, here $s \in \mathcal{P}_{L_a}$); (ii) the BLUPs of SBP, Total Cholesterol, HDL, BMI and smoking (here evaluated not just at $L_a$, but at each time in $\mathcal{P}_{L_a}$); (iii) the cohort under analysis is the landmark cohort (composed of $N_{L_a}$ individuals) in Eq.~\eqref{eq:coxmodel_paige}, while it is the landmark sub-cohort (composed of $N_{L_a,s}$ individuals) in Eq.~\eqref{eq:coxmodel}; (iv) the window of interest (10 years and 5 years respectively).

\begin{equation}
         \label{eq:coxmodel}
\begin{aligned}[b]
        \lambda(t;\textbf{x}_i(s), s) &= 
         \lambda_0(t;s) \cdot \exp\Big\{\textbf{x}_{i,fixed}^T   \boldsymbol{\beta}_{fixed}(s) +
        \textbf{x}_{i,BLUP}^T (s)  \boldsymbol{\beta}_{BLUP}(s)\Big\} \\
        &\quad i \in \{1,..,N_{L_a,s}\} \quad s \in \mathcal{P}_{L_a} \quad s \leq t \leq s + 5.
\end{aligned}
\end{equation}

Given the Nelson-Aalen estimator of the cumulative hazard function $\hat{\Lambda}(t;s)$, $s \leq t \leq s + 5$, and the estimated regression parameters $\hat{\boldsymbol{\beta}}(s)$, $s \in \mathcal{P}_{L_a}$, from Eq.\eqref{eq:coxmodel} and denoting $\textbf{x}_i(s)$ as the vector of all covariates of person $i$ measured at time $s$ and the BLUPs estimated at time $s$, we are able to estimate the 5-year CVD risk  $\hat{r}_i(s + 5;\textbf{x}_i(s), s)$ for person $i$ as follows:
\begin{equation}
\label{eq:risk}
\begin{aligned}[b]
       \hat{r}(s + 5;\textbf{x}_i(s), s) = 1 - \exp\{-\hat{\Lambda}_0(s + 5; s) \cdot \exp\{\textbf{x}_i^T(s) \hat{\boldsymbol{\beta}} (s)\}\} \quad i = \{1,..,N_{L_a,s}\}, \quad s \in \mathcal{P}_{L_a}.
\end{aligned}
\end{equation}

At each landmark age $L_a$, we are able to compute a vector of 5-year CVD risks for each individual $i$ in the landmark cohort. The elements of this vector are the 5-year CVD risk $\hat{r}_i(s + 5;\textbf{x}_i(s), s)$ estimated at each time $s \in \mathcal{P}_{L_a}$.

Finally, we predict the time $t_{i,L_a}^*$ at which the 5-year CVD risk of person $i$ exceeds the 5\% threshold, by linearly interpolating between the first year we estimate a 5-year CVD risk higher than $5\%$ and the previous year.
Note that a person may not cross the risk threshold at all for any $s$ in the prediction time set.

We validate all prediction models using the dynamic concordance index~\citep{harrell1996multivariable, van2011dynamic} and the dynamic Brier Score~\citep{graf1999assessment,van2011dynamic}. Validation was performed using the validation dataset to avoid overfitting. See section 4 of the supplementary material for more details.

\section{Results}
\label{s:results}
\begin{figure}[ht]
    \centering
    \includegraphics[width=\textwidth]{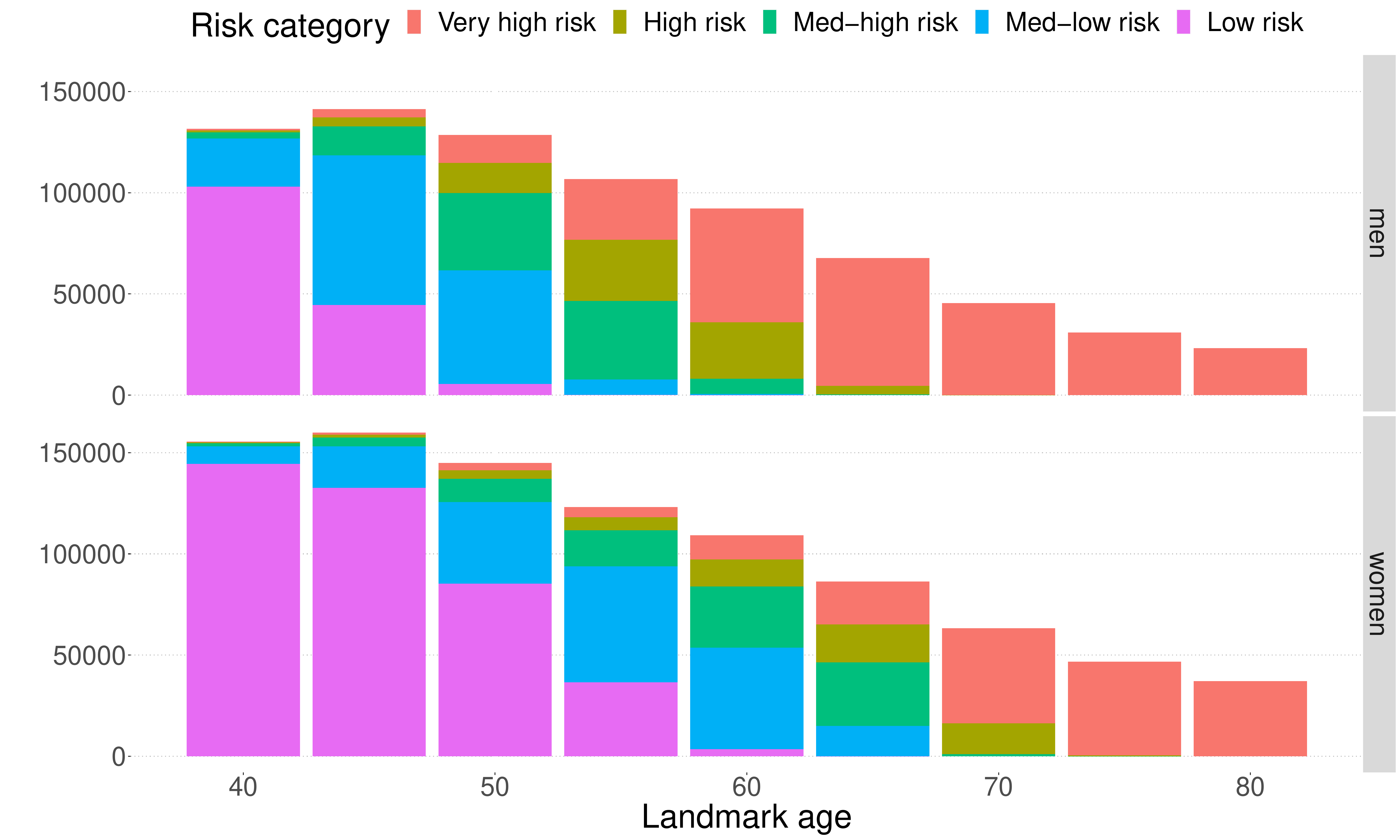}
    \caption{\label{fig:landmark_cohort_size}
    Number of participants in each landmark cohort for men (top row) and women (bottom row) across all landmark ages, in the derivation set. Each color represents the estimated 5-year CVD risk at the landmark age. This figure appears in color in the electronic version of this article.}
\end{figure}

The sizes of the selected landmark cohorts for men (top row) and women (bottom row) are reported in Figure~\ref{fig:landmark_cohort_size}. The colors represent a classification of the 5-year CVD risk at  different landmark ages, i.e., $\hat{r}(L_a + 5; \textbf{x}_i(L_a), L_a)$ from Eq.~\eqref{eq:risk}. The classification is the following: very high if $> 5\%$; high if in the interval $(3.75\%, 5\%]$; medium high if in the interval $(2.5\%, 3.75\%]$; medium low if in the interval $(1.25\%, 2.5\%]$; low if $\leq 1.25\%$.

Note that the biggest landmark cohort size is recorded at landmark age 45 for both women and men. This is not anomalous because people can enter the study after age 40 (see Section~\ref{s:data}). Moreover, as the landmark ages increases, we observe that the proportion of very high risk people increases, while the proportion of low risk people decreases. But note that the sub-cohorts computed at each landmark can only decrease in size.
Following the considerations made at the end of Section~\ref{ss:nb}, we exclude people at very high risk from our risk-assessment strategy evaluation.

\subsection{Optimal risk-assessment strategy}
\label{ss:res_screen}
In this subsection, we present the optimal risk-assessment strategy resulting from Eq.~\eqref{eq:max_INB} for all individuals at each landmark age. 
We set the parameters of Eq.~\eqref{eq:max_INB} as follows: $\lambda =  25,000$ \pounds/year, $u_s =  0.997$, $c_s = 150$ \pounds/year and $c_{\nu} = 18.39$ \pounds/visit.
We evaluated Eq.~\eqref{eq:max_INB} at $F = 10$ different risk-assessment schedules $\boldsymbol{\tau}^{f}$,  $f \in \{1,...,10\}$. A schedule of risk-assessments every 5 years ($f=5$) corresponds to the recommendation of the NICE guidelines.

We represent the result of Net Benefit evaluation at landmark age 40 for women and men in Table~\ref{tab:opt_screen_40}.
We observe that for high risk individuals  more frequent schedules are preferred in general, whereas for people at low or medium-low risk a ten-year schedule appears appropriate.   
The classification as low risk at the landmark age is a good proxy for recommending a 10-year risk-assessment strategy. However, a range of risk-assessment recommendations may be made for individuals with a higher 5-year CVD risk at the landmark age, due to the extra information provided by the values of specific current and future predicted risk factors for those individuals, that is exploited by our prediction model. \\
We can observe in Table~\ref{tab:opt_screen_40} that the greatest part of both cohorts is categorised as low risk (93.09\% women and 78.75\% men), while only 485 (0.31\%) women and 977 (0.75\%) men are labelled as high risk. For almost the whole female cohort (99.56\%) at this landmark age, and for 96.25\% of the male cohort, undergoing visits every 10 years is found to be the optimal configuration. 
Focusing on high risk people, we note that the most recommended risk-assessment strategy is every 1, 2, 3 and 4 years (more evident for men than women). However, there are a few people at high risk whose risk-assessment could be performed every 10 years at landmark age 40 (104 women and 8 men). 
This is because some individuals are predicted flat trends in their 5-year CVD risk profiles. A focus on the 5-year CVD risk profiles for women labelled as high risk at $L_a=40$ is reported in supplementary Figure 3. 

Furthermore, people classified at low risk at $L_a$ are often not expected to initiate statins in the next 10 years. Indeed, looking at supplementary Table 1 and 2, we notice that the 5-year CVD risk is not expected to cross the 5\% threshold for 143,864 of the 144,416 women labelled as low risk at landmark age 40 and for 100,361 of the 102,989 men labelled as low risk at landmark age 40. 

\begin{table}
  \caption{\label{tab:opt_screen_40} Optimal risk-assessment strategy for woman and men at landmark age 40. \\ Women categorised as very high risk are 359 ($0.23\%$) of 155,497, while men at very high risk\\ are  761 ($0.58 \%$) of 131,548.}
  \begin{tabular}{c|c|c|c|c|c|c}
&$f$ &  High risk & Med-high risk & Med-low risk & Low risk &  \\ 
  \hline
\parbox[c]{2mm}{\multirow{8}{*}{\rotatebox[origin=c]{90}{Women}}} & 1 & 76 & 5 & 0 & 0 & 81 (0.05\%) \\ 
   & 2 & 88 & 51 & 4 & 0 & 143 (0.09\%) \\ 
   & 3 & 67 & 65 & 12 & 0 & 144 (0.09\%) \\ 
   & 4 & 5 & 32 & 27 & 0 & 64 (0.04\%) \\ 
   & 5 & 76 & 102 & 0 & 0 & 178 (0.11\%) \\ 
   & 6 & 69 & 6 & 0 & 0 & 75 (0.05\%) \\ 
   & 10 & 104 & 1315 & 8618 & 144416 & 154453 (99.56\%) \\ \cline{2-7}
   & Total & 485 (0.31\%) & 1576 (1.02\%) & 8661 (5.58\%) & 144416 (93.09\%) &  \\ 
   \hline
\parbox[c]{2mm}{\multirow{9}{*}{\rotatebox[origin=c]{90}{Men}}} & 1 & 165 & 3 & 0 & 0 & 168 (0.13\%) \\ 
   & 2 & 288 & 99 & 3 & 0 & 390 (0.3\%) \\ 
   & 3 & 252 & 199 & 10 & 0 & 461 (0.35\%) \\ 
   & 4 & 156 & 363 & 261 & 0 & 780 (0.6\%) \\ 
   & 5 & 58 & 766 & 583 & 0 & 1407 (1.08\%) \\ 
   & 6 & 40 & 637 & 775 & 0 & 1452 (1.11\%) \\ 
   & 7 & 10 & 236 & 0 & 0 & 246 (0.19\%) \\ 
   & 10 & 8 & 767 & 22119 & 102989 & 125883 (96.25\%) \\ \cline{2-7}
   & Total & 977 (0.75\%) & 3070 (2.35\%) & 23751 (18.16\%) & 102989 (78.75\%) &  \\ 
\end{tabular}
\end{table}

\begin{figure}[ht!]
	\centering
	\includegraphics[width =\textwidth]{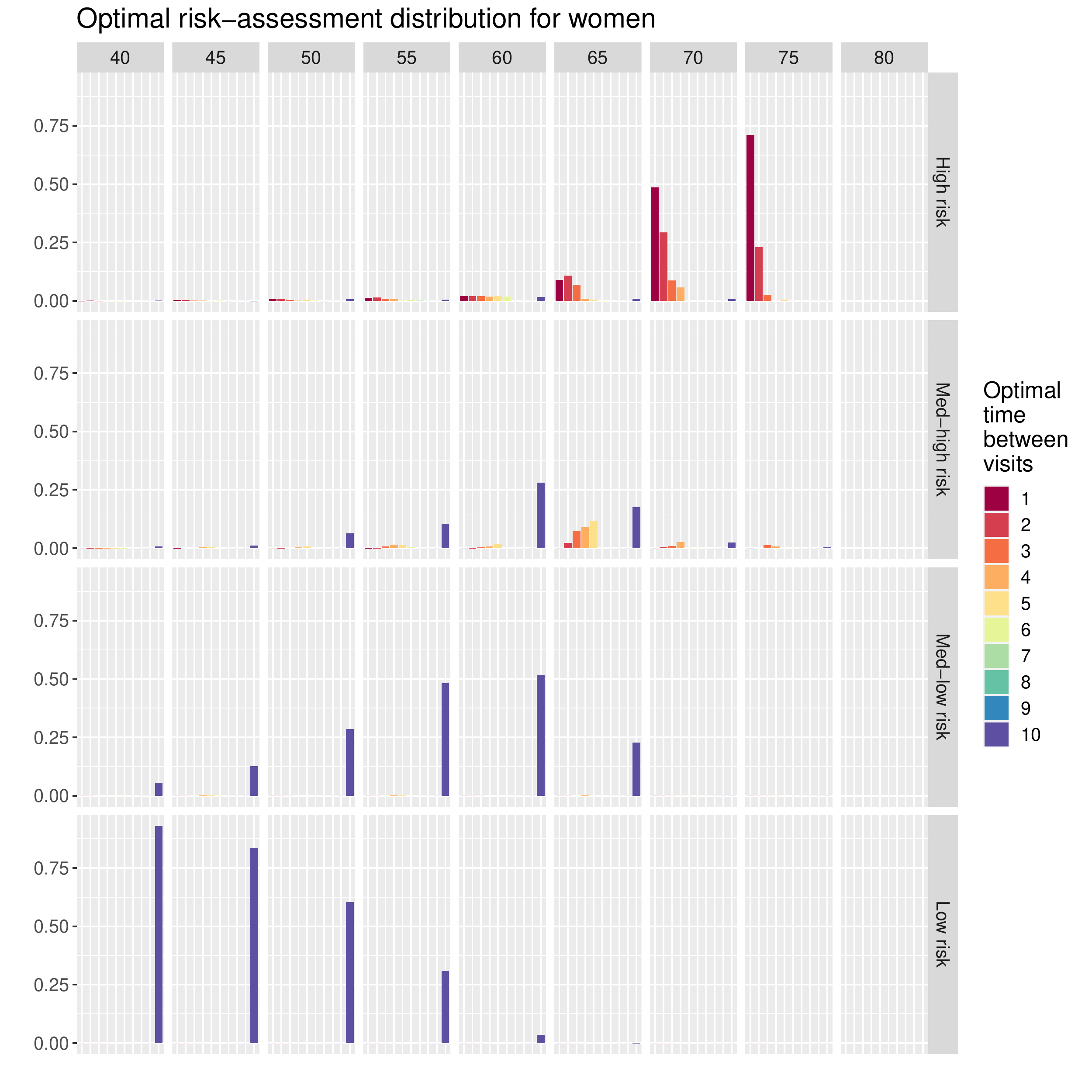}
	\caption{\label{fig:opt_screen_female_all}
Proportions of optimal risk-assessment schedule per each landmark age, for women. This figure appears in color in the electronic version of this article.}
\end{figure}

\begin{figure}[ht!]
	\centering
	\includegraphics[width = \textwidth]{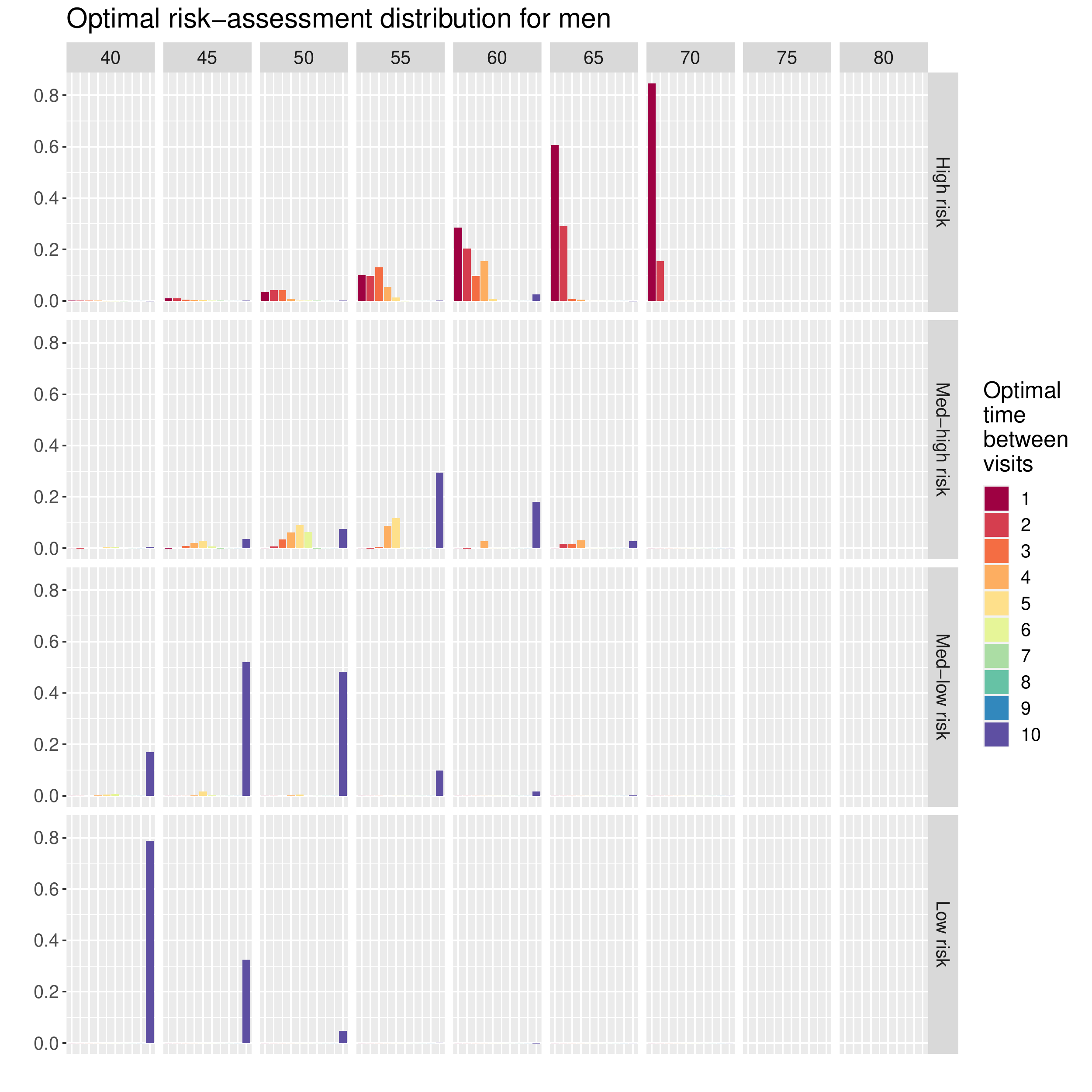}
	\caption{\label{fig:opt_screen_male_all}
Proportions of optimal risk-assessment schedule per each landmark age, for men. This figure appears in color in the electronic version of this article.}
\end{figure}

An overview of the results for women and men across all landmark ages can be found in Figure~\ref{fig:opt_screen_female_all} and Figure~\ref{fig:opt_screen_male_all}. As the landmark age increases, the most frequent optimal risk-assessment strategy shifts from every 10 years to every year for both genders. However, note that for women at landmark age 65 with high  5-year CVD risk at the landmark age the most frequent optimal risk-assessment schedule ranges from every one to three years. Similarly for men from landmark age 55  (Figure~\ref{fig:opt_screen_male_all}). There is a shift of the CVD-risk between men and women. The numbers reported in Figure~\ref{fig:opt_screen_female_all} and~\ref{fig:opt_screen_male_all} are detailed in section 3 of the supplementary material. 

\subsection{Sensitivity analysis
: exploring the effect of NB parameters}
\label{s:sensitivity}

We perform a sensitivity analysis of the NB optimization with respect to the NB parameters $\lambda$, $u_s$, $c_s$ $c_{\nu}$. In general, we observe that results are robust with respect to the parameters choice and minor expected changes are observed. Specifically, $\lambda$ increases, the 10-year frequency is optimal for fewer people, while intermediate frequency becomes optimal for a larger proportion of people. 
A similar observation can be done for the utility associated to statins $u_s$, the lower the impact of statins on the quality of life, the less preferred is the 10-year risk-assessment.  On the contrary, the higher the price of statins, $c_s$, the more risk-assessment strategies associated with less frequent visits are to be preferred.

The complete results of the sensitivity analysis are reported in section 5 of the supplementary material.

\section{Discussion}
\label{s:discussion}
In this paper, we introduced a novel statistical approach to address the multi-faceted problem of identifying optimal risk-assessment strategies for CVD risk prevention. Different CVD risk prevention strategies, such as habit/diet modification and statin prescription, and different risk-assessment schedules have been recommended  worldwide~\citep{ lalor2012guidelines,niceguidelines, pylypchuk2018cardiovascular}. In this work, we focussed on statin initiation because statins have been proven to be the most common CVD prevention method~\citep{reiner2013statins} and we focussed on the UK NICE guidelines~\citep{niceguidelines}. 

The novelty introduced in this work is two-fold: firstly, we provided an extension to the 2-stage landmarking model~\citep{paige2018landmark} in order to estimate the exact time at which the 5-year CVD risk exceeds the 5\% threshold; secondly, we defined a Net Benefit function to discriminate among different visit schemes in order to assess the optimal CVD risk-assessment schedule per person at different landmark ages. 

The extension of the 2-stage landmarking model consisted of  defining a series of landmark sub-cohorts based on a set of prediction times of interest; of estimating BLUPs and of fitting a Cox model based on both fixed covariates and BLUPs,  at each prediction time of interest.

The Net Benefit function is based on the difference between benefits (i.e., CVD free life years) and costs (i.e., quality of life reduction, cost of the visits and of statins purchase) and it is designed as a landmark and person-specific function of the risk-assessment schedule $\boldsymbol{\tau}^f$. The optimal CVD risk-assessment schedule for the $i$-th person ($\boldsymbol{\tau}_{i,L_a}^{opt}$) is the one associated with the highest NB value.

We applied the proposed model to an electronic health record dataset obtained through linking CPRD data to secondary care admissions from HES and mortality records from the ONS. According to our findings, only a portion of the cohort is expected to cross the 5\% threshold and the proportion of this group of people increases with age. Since women have lower CVD incidence than men, more so at younger ages, then assessing CVD risk every 5 years, starting from age 40 for both men and women may be a sub-optimal strategy. Using our method we were able to recommend for each individual at each landmark age the optimal risk-assessment schedule. For lower risk categories with 5 year risk less than 3.75\%, we found that assessing the CVD risk every 10 years is the most frequent optimal choice, while more frequent risk-assessment strategies of every 1 or 2 years were found to be optimal for the majority of the landmark cohort at higher risk. Note that almost all women older than 75 and men older than 70 are labelled as very high risk. This is in line with the fact that age is the most important risk factor for CVD diagnosis.

We had to make some assumptions in order to investigate this complex problem. These assumptions may be limitations of the present study, but also identify directions for further research. For example, we assumed that each person starting statin therapy will be fully compliant, even though statin non-adherence is a well known issue~\citep{simpson2010effects}. Another assumption of our model consisted of censoring deaths both for the identification of the time of crossing the threshold and for the NB computation. This choice is in accordance with the NICE guidelines~\citep{niceguidelines}. Thirdly, we assume a linear trend for the time-varying CVD risk factors, which may not be appropriate for predicting up to 10 years ahead. Finally, we defined a quite general NB function to identify an optimal risk-assessment schedule for a general population. However, the NB function is not able to deal with those people that are labelled as very high risk (5-year CVD risk at a specific landmark age greater than 5\%), and separate recommendations are required for management of CVD risk in this population.  

Future work will further explore these limitations. It is possible to adjust for statin non-adherence by providing a modified $\theta$, or even a time dependent $\theta$. The linear assumption behind the endogenous time-varying variable can be improved by fitting more flexible mixed effects models, although this may require more complete and frequent measurements than are available in the CPRD dataset.  To address the competing risk of death and account for time spent living with CVD, a competing risk or a multi-state model could be defined to assess CVD-specific risk.  A more complex NB function could be designed to take into account both CVD and death. 
Our health outcome included only event-free life years up to 10 years adjusted for quality of life on statins, and we assumed that the cost per QALY gained used by NICE is applicable to these restricted outcomes.
Another possible extension of the NB function could be designed for elder populations, that are completely labelled as very high risk. In this case, the risk-assessment strategy could recommend the \textit{type of measurement} to be taken (i.e., blood tests, SBP,..), instead of the risk-assessment schedule.

\section*{Acknowledgements}
\vspace*{-8pt}
This study is based on data from the Clinical Practice Research Datalink obtained under licence
from the UK Medicines and Healthcare products Regulatory Agency (protocol 162RMn2). The data is provided by patients and
collected by the NHS as part of their care and support. The interpretation and conclusions contained in
this study are those of the author/s alone. F.G. and J.K.B. were funded by the Medical Research Council, unit programme number MRC\_MC\_UU\_00002/5. C.J. was funded by the Medical Research Council, unit programme number MRC\_MC\_UU\_00002/11. 
M.J.S. was funded by the Medical Research Council, the British Heart Foundation, and the National Institute for Health Research’s Blood and Transplant Research Unit (NIHR BTRU) in Donor Health and Genomics (NIHR BTRU-2014-10024). A.M.W. is supported by a British Heart Foundation–Turing Cardiovascular Data Science Award and by the EC-Innovative Medicines Initiative (BigData@Heart).
The work was also supported by the Alan Turing Institute/British Heart Foundation (BHF) (grant SP/18/3/33801). The Cardiovascular Epidemiology Unit is underpinned by core funding from the UK Medical Research Council (MR/L003120/1), British Heart Foundation (RG/13/13/30194 and RG/18/13/33946), and National Institute for Health Research Cambridge Biomedical Research Centre (BRC-1215-20014). For the purpose of open access, the author has applied a Creative Commons Attribution (CC BY) licence to any Author Accepted Manuscript version arising from this submission. 
 
\section*{Supplementary material}
Supplementary material is available in the online version of the article at the publisher’s website. This study is based on data from the Clinical Practice Research Datalink (CPRD) obtained under license from the UK Medicines and Healthcare products Regulatory Agency (protocol 162RMn2). This work uses data provided by patients and collected by the NHS as part of their care and support.  Code is publicly available at \url{https://github.com/fgaspe04/CPRD/}. 

\bibliographystyle{plainnat}  
\bibliography{biblio}

\end{document}


\maketitle

\section{Cohort selection, risk factors and outcome definitions}
In this section we report details related to cohort selection and variables included in the proposed model.

\begin{figure}[ht!]
    \centering
    \includegraphics[width = 0.8\textwidth]{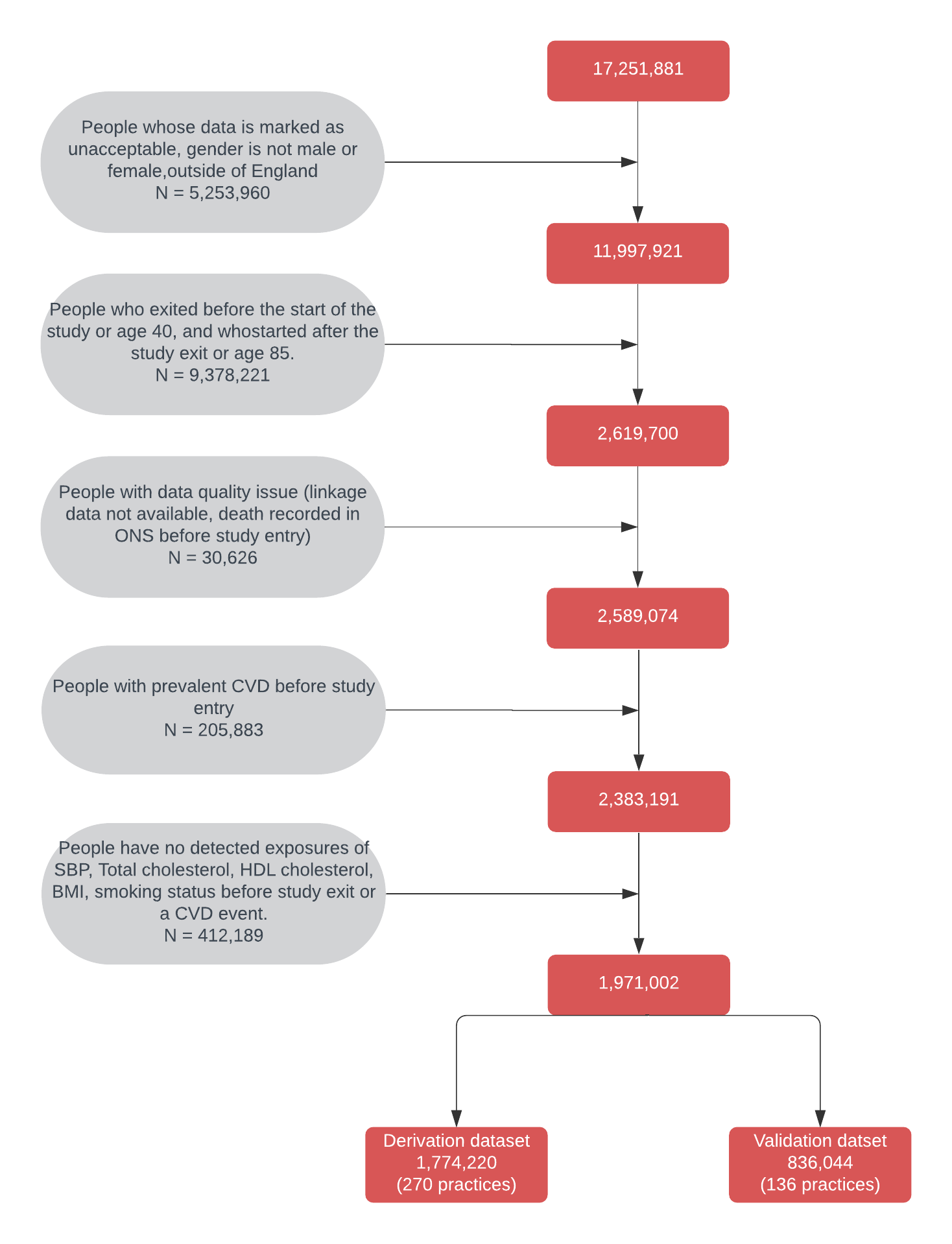}
    \caption{Flow chart of the selection process of the analysed cohort.}
    \label{fig:flowchart}
\end{figure}

In Figure~\ref{fig:flowchart}, we represent the scheme of the cohort selection. The final derivation dataset is composed of 1,774,220 people distributed across 270 practices in the UK; while the validation dataset is composed of 836,044 people distributed in 136 practices. 

Following \citet{xu2021prediction}, we define the \textit{study entry} for each person as the latest of the following four dates: the date of 6 months after registration at the general practice; the date the individual turned 30 years of age; the date that the data for the practice were up to standard \citep{tate2017quality}; or April 01, 2004, the date that the Quality and Outcomes Framework (QOF) was introduced \citep{nhsdigital}. We define the \textit{study exit} for each person as the earliest of the following dates: the date of deregistration at the practice; the individual’s death; the date that the individual turned 95 years of age; the last contact date for the practice with CPRD; or the administration end date (November 2017). 

The Read codes (used to identify outcomes in CPRD) and International Classification of Diseases, Tenth Revision, codes (used to identify outcomes in primary or secondary diagnosis fields from Hospital Episode Statistics and in underlying or subordinate cause of death fields from the Office for National Statistics) are provided in the Web Appendix 1, Web Tables 1 and 2 of~\citet{xu2021prediction}.

Previous diagnosis of diabetes, renal disease, depression, migraine, severe mental illness, rheumatoid arthritis and atrial fibrillation are ascertained from CPRD Read codes.
 Blood pressure medication (yes/no) is ascertained from CPRD prescription information and it is defined as the date of first prescription.
 Statin initiation is defined as the date of first CPRD prescription (code list for CPRD prescription provided in Web Appendix 2, Web Table 3 of~\citet{xu2021prediction}).
 Finally, Townsend deprivation index ranges from 1 to 20. This index presents a total of 1979 missing values (0.08\% of the whole cohort), that are imputed through the mean value per each landmark.
 
 We set to missing biologically implausible values: BMI $>80$ kg/m$^2$ ; SBP $>250$ mmHg or $<60$ mmHg; total cholesterol level $>20$ mmol/L or $<1.75$ mmol/L; HDL cholesterol level $>3.1$ mmol/L or $<0.3$ mmol/L.

 
Furthermore, we consider two different sets of risk factors, if we are performing the analysis before or after landmark age $60$.
At all landmark ages, we include the BLUPs of BMI, HDL, SBP, total cholesterol and smoking. At all landmark ages, we include blood pressure medication, Townsend deprivation index, previous diagnosis of diabetes, depression, migraine and severe mental illness. Previous diagnosis of renal disease, rheumatoid arthritis and atrial fibrillation are included only after landmark age $60$. This choice is motivated by the fact that these specific conditions are extremely rare at younger ages and the estimates of Cox model in Eq. 12 of the main manuscript are unfeasible.

\section{Details of the Incremental Net Benefit function}
\subsection{Derivation of EFLY before and after statins initiation}
We evaluate as benefit the restricted event free life years (EFLY), i.e. we investigate the time to CVD diagnosis, $T$, restricted to the observable time window $[L_a, L_a + 10]$. 

To define the restricted EFLY, $min(T, L_a +10)$, we assume that the time to CVD can be written as $T = T^{NS} + T^S$, where $T^{NS}$ is event-free time elapsed before statin initiation and $T^{S}$ is the event-free time elapsed after statin initiation. 
The definition of the distribution of the time to CVD, $T$, should reflect the fact that statin usage has a positive impact on CVD-free life expectancy \citep{ferket2012personalized}. In order to express this gain in EFLY, we quantify the effect of statins via the hazard ratio $\theta$. In this article, we set $\theta = 0.8$, following previous meta-analysis of statin trials~\citep{unit2005efficacy}. 

The discontinuity point in the definition of $T$, that separates the time to CVD without statins $T^{NS}$ and the time with statins $T^S$, is defined as $\tau_{k^*_{i,L_a}}$, the first visit among the scheduled ones $\boldsymbol{\tau}$ that happens after $t^*_{i,L_a}$, the predicted time when the 5-year CVD risk of person $i$ exceeds the 5\% threshold at landmark age $L_a$. $\tau_{k^*_{i,L_a}}$ is landmark and person-specific and depends on the risk-assessment strategy under evaluation, $\boldsymbol{\tau}$. 

We can define the hazard rate of CVD onset, given the personal covariates known at time $L_a$ and $\tau_{k^*_i}$ as reported in Eq.~\eqref{eq:density}.

\begin{flalign}
    \lambda(t;\textbf{x}_i(L_a), L_a, \tau_{k^*_{i,L_a}}) &= \lambda^{NS}(t;\textbf{x}_i(L_a), L_a)  \cdot \mathbbm{1}{ \{t \leq \tau_{k^*_{i,L_a}}\} } + \lambda^{S}(t;\textbf{x}_i(L_a), L_a) \cdot \mathbbm{1}{ \{t > \tau_{k^*_{i,L_a}}\} }   \nonumber \\[3pt]
     &= \lambda_0(t;L_a) \cdot \exp\{\textbf{x}_i(L_a)^T \boldsymbol{\beta}(L_a)\} \cdot \left[ \mathbbm{1}{ \{t \leq \tau_{k^*_{i,L_a}}\} } + \theta \cdot \mathbbm{1}{ \{t > \tau_{k^*_{i,L_a}}\} } \right],   \quad t\geq L_a.
     \label{eq:density}
\end{flalign}
We can compute the cumulative hazard function $\Lambda(t;\textbf{x}_i(L_a), L_a, \tau_{k^*_{i,L_a}})$ as explained in Eq.~\eqref{eq:cumlambda}.

\begin{flalign}
    \Lambda(t;\textbf{x}_i(L_a), L_a, \tau_{k^*_{i,L_a}}) &= \int_{L_a}^t\lambda_0(u) \cdot \exp\{\textbf{x}_i(L_a)^T \boldsymbol{\beta}(L_a)\} \cdot \left[ \mathbbm{1}{ \{u \leq \tau_{k^*_{i,L_a}}\} } + \theta \cdot \mathbbm{1}{ \{u > \tau_{k^*_{i,L_a}}\} } \right] \, du  \nonumber \\[3pt]
    &  = \begin{cases}
    \Lambda_0(t;L_a)\exp\{\textbf{x}_i(L_a)^T \boldsymbol{\beta}(L_a)\},  &t \leq \tau_{k^*_{i,L_a}} \\
    \Lambda_0(\tau_{k^*_{i,L_a}})\exp\{\textbf{x}_i(L_a)^T \boldsymbol{\beta}(L_a)\}  + (\Lambda_0(t;L_a) - \Lambda_0(\tau_{k^*_{i,L_a}})) \cdot \theta \cdot  \exp\{\textbf{x}_i(L_a)^T \boldsymbol{\beta}(L_a)\}  ,  &t > \tau_{k^*_{i,L_a}}
    \end{cases}
     \label{eq:cumlambda}
\end{flalign}
Defining $\Lambda^{NS}(t;\textbf{x}_i(L_a), L_a)$ as $\Lambda_0(t;L_a)\exp\{\textbf{x}_i(L_a)^T \boldsymbol{\beta}(L_a)\}$, we can rewrite Eq.~\eqref{eq:cumlambda} as follows:
\begin{flalign}
\Lambda(t;\textbf{x}_i(L_a), L_a, \tau_{k^*_{i,L_a}}) &= 
    \begin{cases}
    \Lambda^{NS}(t;\textbf{x}_i(L_a), L_a),  &t \leq \tau_{k^*_{i,L_a}} \\
    \Lambda^{NS}(\tau_{k^*_{i,L_a}};\textbf{x}_i(L_a), L_a)  + (\Lambda^{NS}(t;\textbf{x}_i(L_a), L_a) - \Lambda^{NS}(\tau_{k^*_{i,L_a}};\textbf{x}_i(L_a), L_a)) \cdot \theta,  &t > \tau_{k^*_{i,L_a}}
    \end{cases}
     \label{eq:cumlambda2}
\end{flalign}
Finally, we the survival function $S(t;\textbf{x}_i(L_a), \tau_{k^*_{i,L_a}})$ in Eq.~\eqref{eq:survival}.
\begin{flalign}
    S(t;\textbf{x}_i(L_a), L_a, \tau_{k^*_{i,L_a}}) &=  \exp\{-\Lambda(t;\textbf{x}_i(L_a), \tau_{k^*_{i,L_a}}, L_a)\} = \begin{cases}
    S^{NS}(t;\textbf{x}_i(L_a), L_a),  &t \leq \tau_{k^*_{i,L_a}} \\
    S^{S} (t;\textbf{x}_i(L_a), L_a),  &t > \tau_{k^*_{i,L_a}}
    \end{cases}
     \label{eq:survival}
\end{flalign}
where:
\begin{equation}
    S^{S} (t;\textbf{x}_i(L_a), L_a) = S^{NS}(\tau_{k^*_{i,L_a}};\textbf{x}_i(L_a), L_a) \cdot \left( \frac{S^{NS}(t;\textbf{x}_i(L_a), L_a)}{ S^{NS}(\tau_{k^*_{i,L_a}};\textbf{x}_i(L_a), L_a)} \right)^{\theta}
    \label{eq:s_s}
\end{equation}

Note that if a person is never prescribed statins, then we have $S(t;\textbf{x}_i(L_a), L_a) = S^{NS}(t;\textbf{x}_i(L_a), L_a)$. 

Conditioning on $\tau_{k^*_{i,L_a}}$, $\tau_{k^*_{i,L_a}} \leq L_a + 10$, we are able to define the restricted EFLY in Eq.~\eqref{eq:expectedrestr}.  

\begin{flalign}
    EFLY &= EFLY_{NS}(\tau_{k^*_{i,L_a}}) + EFLY_S(\tau_{k^*_{i,L_a}}) \nonumber \\
    &= \int_{L_a}^{\tau_{k^*_{i,L_a}}} S^{NS}(t;\textbf{x}_i(L_a), L_a) \, dt + \int_{\tau_{k^*_{i,L_a}}}^{L_a + 10} S^{S}(t;\textbf{x}_i(L_a), L_a) \, dt.
    \label{eq:expectedrestr}
\end{flalign}

\subsection{Expected number of risk assessments}
The expected number of risk assessment, $\mathbb{E}_{\boldsymbol{\tau}}[N_i]$, is part of the costs associated with a specific risk assesment strategy $\boldsymbol{\tau}$. We assume that the CVD-risk assessment of a person is performed up to time $\tau_{k^*_{i,L_a}}$ (i.e., no more visits after statins initiation).

\begin{figure}
    \centering
    \includegraphics[trim = {8cm 2cm 8cm 5cm},width = 8cm]{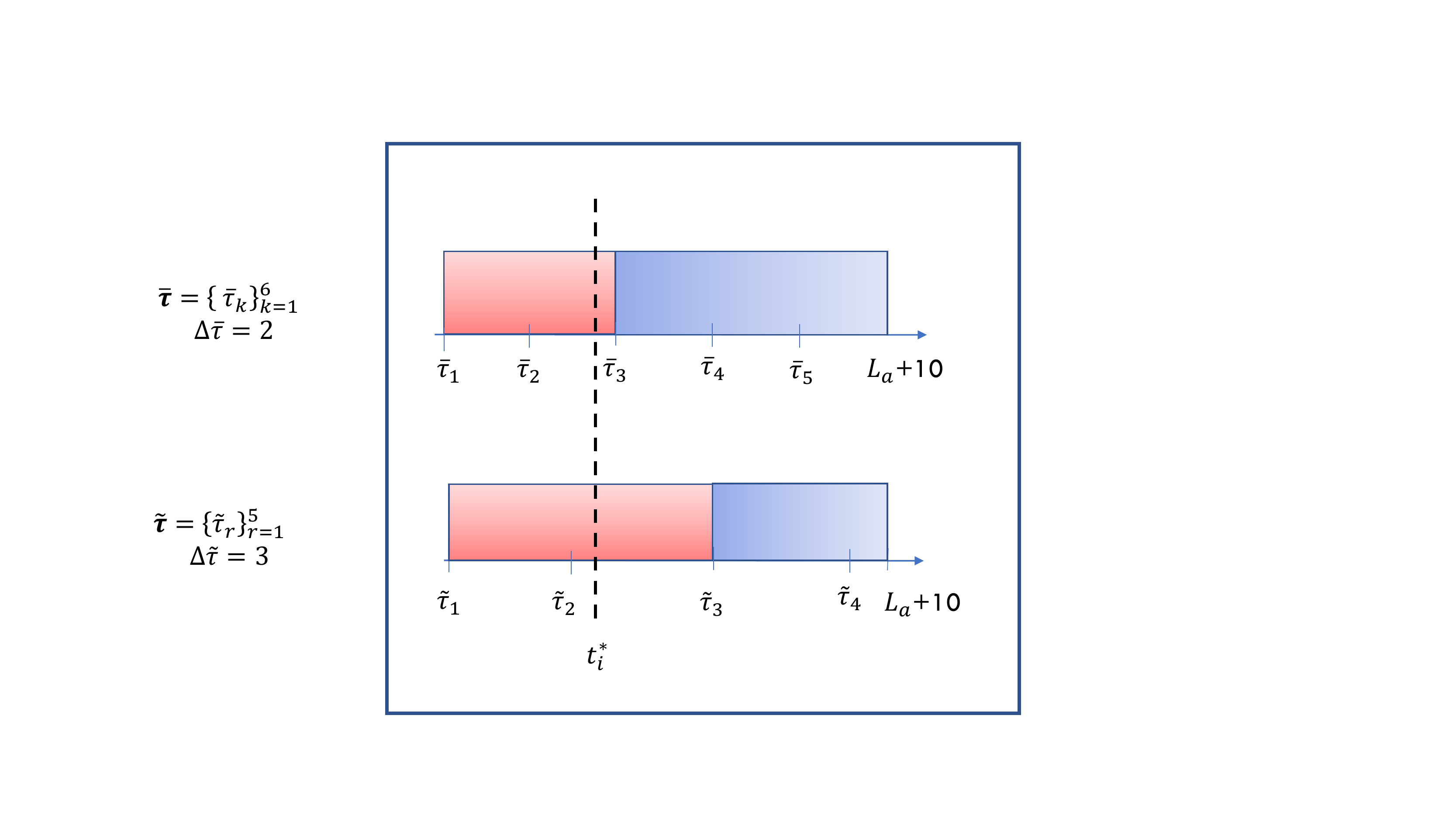} 
    \caption{Example figure for describing the procedure for computing $k^*_i$, $\tau_{k^*_i}$, given $t_i^*$ and two specific risk-assessment strategies. In the upper part of the figure, we consider risk-assessment schedule $\bar{\boldsymbol{\tau}}$ (visits every 2 years), while in the lower part, we consider another risk-assessment schedule $\tilde{\boldsymbol{\tau}}$ (visits every 3 years). This  person is expected to cross the $5\%$ threshold at $t_i^*$, which implies that this person is going to start taking statins at the third visit ($\bar{k}_i^* = \tilde{k}_i^* = 3$) in both cases ($\mathbb{E}_{\bar{\boldsymbol{\tau}}}[N_i] =  \mathbb{E}_{\tilde{\boldsymbol{\tau}}}[N_i] = 3$). However, $\bar{\tau}_{3}$ and $\tilde{\tau}_{3}$ are different, $\bar{\tau}_{3} < \tilde{\tau}_{3}$, which means that time spent under statins blue boxes) is longer if we focus on $\bar{\boldsymbol{\tau}}$ (risk-assessment every 2 years). The statin-free time is represented through red boxes.}
    \label{fig:example_nb}
\end{figure}

In Fig.~\ref{fig:example_nb}, we represent an illustrative example to show how the expected number of visits is computed for two different risk-assessment schedules ($\bar{\boldsymbol{\tau}}$ in the top row and $\tilde{\boldsymbol{\tau}}$ in the bottom row) for a generic person whose 5-year CVD risk exceeds the 5\% threshold at $t^*_i$ (dashed black line).
According to both risk-assessment schedules, person $i$ should start taking statins from the third visit, which means $\bar{k}^*_i =\tilde{k}^*_i = 3$ and $\mathbb{E}_{\bar{\boldsymbol{\tau}}}[N_i] =  \mathbb{E}_{\tilde{\boldsymbol{\tau}}}[N_i] = 3$.

If a person never crosses the $5\%$ threshold, they will never start taking statins and the expected number of visits including the baseline visit is $\mathbb{E}_{\boldsymbol{\tau}}[N_i] = 1+10/\Delta \tau$, where $\Delta \tau$ is the time between visits according to visit schedule $\boldsymbol{\tau}$. 
From Fig. \ref{fig:example_nb}, the expected number of visits for a person whose 5-year CVD risk never crosses the 5\% threshold are 6, according to $\bar{\boldsymbol{\tau}}$ and 4.33, according to $\tilde{\boldsymbol{\tau}}$.

\section{Optimal risk-assessment scheduling: CVD free life years}
\label{sec:cvdotherlmage}
In this section, we report the recommended risk-assessment strategies across different landmark ages for women in Table~\ref{tab:overall_CVD_female} and for men in Table~\ref{tab:overall_CVD_male}. In each table, we describe the results based on the landmark ages (values in columns) and on the 5-year CVD risk categories estimated at each landmark age (values in rows). The percentage associated to each number of the table is computed with respect to the landmark cohort. For example, if we focus on women at high risk at landmark age 40 (third column from the left in Table~\ref{tab:overall_CVD_female}), we note that 76 (0.05\% of 155497) women are recommended to have a risk-assessment every year.
Furthermore, we report the total number of people belonging to each risk category and the total number of people among them, whose 5-year CVD risk is not expected to cross the 5\% threshold in the next 10 years.
These two numbers are at the bottom of each risk category block.
For example, if we focus on women at high risk at landmark age 40 (third column from the left in Table~\ref{tab:overall_CVD_female}), we note that 485 are labelled as High risk and 50 (10.31\%) of them are not expected to cross the 5\% threshold.
At the bottom of both tables we record the landmark cohort size and total number of people whose 5-year CVD risk is not expected to cross the 5\% threshold in the next 10 years.  
Looking at these rows, we note that the landmark cohort size decreases over landmark ages. The total number of people whose 5\% CVD risk is not expected to cross the threshold decreases over landmark ages. This line should be read in pair with the top one where the total number of people labelled as very high risk is reported (note that these numbers increase over landmark ages). These observations hold for both women and men. 

Relevant differences between women in Table~\ref{tab:overall_CVD_female} and men in Table~\ref{tab:overall_CVD_male} are related to the time when the number of people whose 5-year CVD risk is not expected to cross the threshold starts dropping (last row) and the time when the percentage of people labelled as very high risk starts increasing (top row). Indeed, for women the first number start dropping at landmark age 65 (from 34.79\% to 8.18\%), while for men at landmark age 55 (from 43.01\% to 5.6\%). Analogously, almost a quarter of the age 65 landmark cohort of women (24.62\%) is labelled as very high risk, similar number is reached by men already at landmark age 55 (28.1\%).

\begin{table}[ht!]
\centering
\caption{\\Optimal CVD-risk assessment frequency across landmark ages (in columns), stratified by baseline risk category (in rows). This table refers to women and the outcome of interest is 10-year CVD. In the two bottom lines we report the landmark cohort size and the total number of people whose 5-year CVD risk is not expected to cross the 5\% threshold in the next 10 years.}
\label{tab:overall_CVD_female}
\resizebox{\textwidth}{!}{%
\begin{tabular}{ccccccccccc}
  \hline \hline
Risk class & Optimal  & 40 & 45 & 50 & 55 & 60 & 65 & 70 & 75 & 80 \\ 
  \midrule
Very high  &  - & 359 (0.23\%) & 1119 (0.7\%) & 3586 (2.47\%) & 5068 (4.12\%) & 11953 (10.95\%) & 21274 (24.62\%) & 46926 (74.18\%) & 46361 (99.09\%) & 37178 (100\%) \\ 
  \midrule
  \parbox[t]{10mm}{\multirow{12}{*}{\rotatebox[origin=c]{90}{High}}}  &  1 & 76 (0.05\%) & 405 (0.25\%) & 876 (0.6\%) & 1498 (1.22\%) & 1995 (1.83\%) & 5795 (6.71\%) & 7928 (12.53\%) & 303 (0.65\%) & - \\ 
   &  2 & 88 (0.06\%) & 465 (0.29\%) & 969 (0.67\%) & 1746 (1.42\%) & 1963 (1.8\%) & 7044 (8.15\%) & 4798 (7.58\%) & 98 (0.21\%) & - \\ 
   &  3 & 67 (0.04\%) & 96 (0.06\%) & 585 (0.4\%) & 1004 (0.82\%) & 2000 (1.83\%) & 4434 (5.13\%) & 1431 (2.26\%) & 11 (0.02\%) & - \\ 
   &  4 & 5 (0\%) & 144 (0.09\%) & 163 (0.11\%) & 876 (0.71\%) & 1759 (1.61\%) & 463 (0.54\%) & 946 (1.5\%) & - & - \\ 
   &  5 & 76 (0.05\%) & 69 (0.04\%) & 387 (0.27\%) & 217 (0.18\%) & 2108 (1.93\%) & 351 (0.41\%) & - & 2 (0\%) & - \\ 
   &  6 & 69 (0.04\%) & 128 (0.08\%) & 263 (0.18\%) & 491 (0.4\%) & 1871 (1.71\%) & 10 (0.01\%) & - & - & - \\ 
   &  7 & - & 7 (0\%) & 17 (0.01\%) & 99 (0.08\%) & - & - & - & - & - \\ 
   &  8 & - & - & - & - & - & - & - & - & - \\ 
   &  9 & - & - & - & - & - & - & - & - & - \\ 
   &  10 & 104 (0.07\%) & 84 (0.05\%) & 1000 (0.69\%) & 522 (0.42\%) & 1632 (1.49\%) & 617 (0.71\%) & 113 (0.18\%) & - & - \\ \cline{2-11}
   &  Total & 485 & 1398 & 4260 & 6453 & 13328 & 18714 & 15216 & 414 & 0 \\ 
   & Never cross & 50 (10.31\%) & 8 (0.57\%) & 438 (10.28\%) & 82 (1.27\%) & 3 (0.02\%) & 123 (0.66\%) & 2 (0.01\%) & - & - \\ 
  \midrule
  \parbox[t]{10mm}{\multirow{12}{*}{\rotatebox[origin=c]{90}{Med-high}}}  &  1 & 5 (0\%) & 6 (0\%) & 1 (0\%) & 10 (0.01\%) & - & - & - & - & - \\ 
   &  2 & 51 (0.03\%) & 235 (0.15\%) & 118 (0.08\%) & 123 (0.1\%) & 58 (0.05\%) & 1432 (1.66\%) & 111 (0.18\%) & 1 (0\%) & - \\ 
   &  3 & 65 (0.04\%) & 448 (0.28\%) & 210 (0.14\%) & 950 (0.77\%) & 342 (0.31\%) & 4915 (5.69\%) & 151 (0.24\%) & 6 (0.01\%) & - \\ 
   &  4 & 32 (0.02\%) & 762 (0.48\%) & 529 (0.37\%) & 1827 (1.48\%) & 772 (0.71\%) & 5868 (6.79\%) & 445 (0.7\%) & 3 (0.01\%) & - \\ 
   &  5 & 102 (0.07\%) & 543 (0.34\%) & 1178 (0.81\%) & 1651 (1.34\%) & 1864 (1.71\%) & 7755 (8.98\%) & - & - & - \\ 
   &  6 & 6 (0\%) & 382 (0.24\%) & 433 (0.3\%) & 807 (0.66\%) & 2 (0\%) & - & - & - & - \\ 
   &  7 & - & - & - & - & - & - & - & - & - \\ 
   &  8 & - & - & - & - & - & - & - & - & - \\ 
   &  9 & - & - & - & - & - & - & - & - & - \\ 
   &  10 & 1315 (0.85\%) & 1911 (1.19\%) & 8943 (6.17\%) & 12354 (10.03\%) & 27238 (24.95\%) & 11430 (13.23\%) & 408 (0.64\%) & 2 (0\%) & - \\ \cline{2-11}
   &  Total & 1576 & 4287 & 11412 & 17722 & 30276 & 31400 & 1115 & 12 & 0 \\ 
   & Never cross & 821 (52.09\%) & 627 (14.63\%) & 4870 (42.67\%) & 7729 (43.61\%) & 1138 (3.76\%) & 1437 (4.58\%) & 2 (0.18\%) & - & - \\ 
  \midrule
  \parbox[t]{10mm}{\multirow{12}{*}{\rotatebox[origin=c]{90}{Med-Low}}}  &  1 & - & - & - & - & - & - & - & - & - \\ 
   &  2 & 4 (0\%) & - & - & - & - & - & - & - & - \\ 
   &  3 & 12 (0.01\%) & 52 (0.03\%) & 1 (0\%) & 35 (0.03\%) & - & 14 (0.02\%) & - & - & - \\ 
   &  4 & 27 (0.02\%) & 237 (0.15\%) & 20 (0.01\%) & 177 (0.14\%) & 12 (0.01\%) & 123 (0.14\%) & - & - & - \\ 
   &  5 & - & 174 (0.11\%) & 12 (0.01\%) & 132 (0.11\%) & - & - & - & - & - \\ 
   &  6 & - & - & - & - & - & - & - & - & - \\ 
   &  7 & - & - & - & - & - & - & - & - & - \\ 
   &  8 & - & - & - & - & - & - & - & - & - \\ 
   &  9 & - & - & - & - & - & - & - & - & - \\ 
   &  10 & 8618 (5.54\%) & 20071 (12.55\%) & 40276 (27.79\%) & 56987 (46.28\%) & 50104 (45.89\%) & 14861 (17.2\%) & - & - & - \\ \cline{2-11}
   &  Total & 8661 & 20534 & 40309 & 57331 & 50116 & 14998 & 0 & 0 & 0 \\ 
   & Never cross & 7692 (88.81\%) & 16253 (79.15\%) & 32028 (79.46\%) & 54396 (94.88\%) & 33458 (66.76\%) & 5493 (36.62\%) & - & - & - \\ 
  \midrule
  \parbox[t]{10mm}{\multirow{12}{*}{\rotatebox[origin=c]{90}{Low}}} &  1 & - & - & - & - & - & - & - & - & - \\ 
   &  2 & - & - & - & - & - & - & - & - & - \\ 
   &  3 & - & - & - & - & - & - & - & - & - \\ 
   &  4 & - & - & - & - & - & - & - & - & - \\ 
   &  5 & - & - & - & - & - & - & - & - & - \\ 
   &  6 & - & - & - & - & - & - & - & - & - \\ 
   &  7 & - & - & - & - & - & - & - & - & - \\ 
   &  8 & - & - & - & - & - & - & - & - & - \\ 
   &  9 & - & - & - & - & - & - & - & - & - \\ 
   &  10 & 144416 (92.87\%) & 132637 (82.91\%) & 85340 (58.89\%) & 36561 (29.69\%) & 3500 (3.21\%) & 17 (0.02\%) & - & - & - \\ 
   \cline{2-11}
   &  Total & 144416 & 132637 & 85340 & 36561 & 3500 & 17 & 0 & 0 & 0 \\ 
   & Never cross & 143864 (99.62\%) & 131791 (99.36\%) & 81119 (95.05\%) & 36540 (99.94\%) & 3384 (96.69\%) & 17 (100\%) & - & - & - \\ 
   \midrule
   & Total & 155497 & 159975 & 144907 & 123135 & 109173 & 86403 & 63257 & 46787 & 37178 \\ 
   & Never cross & 152427 (98.03\%) & 148679 (92.94\%) & 118455 (81.75\%) & 98747 (80.19\%) & 37983 (34.79\%) & 7070 (8.18\%) & 4 (0.01\%) & - & - \\ 
   \hline
\end{tabular}}
\end{table}

\begin{table}[ht!]
\centering
\caption{\\Optimal CVD-risk assessment frequency across landmark ages (in columns), stratified by baseline risk category (in rows). This table refers to men and the outcome of interest is 10-year CVD. In the two bottom lines we report the landmark cohort size and the total number of people whose 5-year CVD risk is not expected to cross the 5\% threshold in the next 10 years.}
\label{tab:overall_CVD_male}
\resizebox{\textwidth}{!}{%
\begin{tabular}{ccccccccccc}
  \toprule
Risk class & Optimal  & 40 & 45 & 50 & 55 & 60 & 65 & 70 & 75 & 80 \\ 
  \midrule
Very high  &  - & 761 (0.58\%) & 4077 (2.89\%) & 13792 (10.73\%) & 30009 (28.1\%) & 56178 (60.93\%) & 63144 (93.15\%) & 45486 (99.94\%) & 30978 (100\%) & 23189 (100\%) \\ 
  \midrule
  \parbox[t]{10mm}{\multirow{12}{*}{\rotatebox[origin=c]{90}{High}}}  &  1 & 165 (0.13\%) & 1385 (0.98\%) & 3956 (3.08\%) & 7634 (7.15\%) & 10305 (11.18\%) & 2816 (4.15\%) & 22 (0.05\%) & - & - \\ 
   &  2 & 288 (0.22\%) & 1388 (0.98\%) & 4891 (3.8\%) & 7441 (6.97\%) & 7357 (7.98\%) & 1350 (1.99\%) & 4 (0.01\%) & - & - \\ 
   &  3 & 252 (0.19\%) & 606 (0.43\%) & 4893 (3.81\%) & 9968 (9.33\%) & 3500 (3.8\%) & 32 (0.05\%) & - & - & - \\ 
   &  4 & 156 (0.12\%) & 331 (0.23\%) & 669 (0.52\%) & 4094 (3.83\%) & 5584 (6.06\%) & 24 (0.04\%) & - & - & - \\ 
   &  5 & 58 (0.04\%) & 386 (0.27\%) & 161 (0.13\%) & 968 (0.91\%) & 211 (0.23\%) & - & - & - & - \\ 
   &  6 & 40 (0.03\%) & 82 (0.06\%) & 205 (0.16\%) & 9 (0.01\%) & - & - & - & - & - \\ 
   &  7 & 10 (0.01\%) & 105 (0.07\%) & 85 (0.07\%) & - & - & - & - & - & - \\ 
   &  8 & - & - & - & - & - & - & - & - & - \\ 
   &  9 & - & - & - & - & - & - & - & - & - \\ 
   &  10 & 8 (0.01\%) & 222 (0.16\%) & 98 (0.08\%) & 117 (0.11\%) & 909 (0.99\%) & 2 (0\%) & - & - & - \\ \cline{2-11}
   &  Total & 977 & 4505 & 14958 & 30231 & 27866 & 4224 & 26 & 0 & 0 \\ 
   & Never cross & 1 (0.1\%) & 139 (3.09\%) & 70 (0.47\%) & 12 (0.04\%) & 61 (0.22\%) & - & - & - & - \\ 
  \midrule
  \parbox[t]{10mm}{\multirow{12}{*}{\rotatebox[origin=c]{90}{Med-high}}}   &  1 & 3 (0\%) & 24 (0.02\%) & - & 1 (0\%) & - & - & - & - & - \\ 
   &  2 & 99 (0.08\%) & 228 (0.16\%) & 792 (0.62\%) & 60 (0.06\%) & 9 (0.01\%) & 81 (0.12\%) & - & - & - \\ 
   &  3 & 199 (0.15\%) & 1211 (0.86\%) & 4038 (3.14\%) & 418 (0.39\%) & 76 (0.08\%) & 70 (0.1\%) & - & - & - \\ 
   &  4 & 363 (0.28\%) & 2820 (2\%) & 7097 (5.52\%) & 6711 (6.28\%) & 991 (1.07\%) & 144 (0.21\%) & - & - & - \\ 
   &  5 & 766 (0.58\%) & 3935 (2.78\%) & 10357 (8.06\%) & 9012 (8.44\%) & - & - & - & - & - \\ 
   &  6 & 637 (0.48\%) & 962 (0.68\%) & 7175 (5.58\%) & - & - & - & - & - & - \\ 
   &  7 & 236 (0.18\%) & 150 (0.11\%) & 50 (0.04\%) & - & - & - & - & - & - \\ 
   &  8 & - & - & - & - & - & - & - & - & - \\ 
   &  9 & - & - & - & - & - & - & - & - & - \\ 
   &  10 & 767 (0.58\%) & 4954 (3.51\%) & 8638 (6.72\%) & 22640 (21.2\%) & 6485 (7.03\%) & 125 (0.18\%) & - & - & - \\ \cline{2-11}
   &  Total & 3070 & 14284 & 38147 & 38842 & 7561 & 420 & 0 & 0 & 0 \\ 
   & Never cross & 116 (3.78\%) & 1839 (12.87\%) & 2287 (6\%) & 2258 (5.81\%) & 937 (12.39\%) & - & - & - & - \\ 
  \midrule
  \parbox[t]{10mm}{\multirow{12}{*}{\rotatebox[origin=c]{90}{Medium-low}}}  &  1 & - & - & - & - & - & - & - & - & - \\ 
   &  2 & 3 (0\%) & - & - & - & - & - & - & - & - \\ 
   &  3 & 10 (0.01\%) & - & 13 (0.01\%) & 1 (0\%) & - & - & - & - & - \\ 
   &  4 & 261 (0.2\%) & 132 (0.09\%) & 221 (0.17\%) & 8 (0.01\%) & - & - & - & - & - \\ 
   &  5 & 583 (0.44\%) & 2191 (1.55\%) & 501 (0.39\%) & 1 (0\%) & - & - & - & - & - \\ 
   &  6 & 775 (0.59\%) & 237 (0.17\%) & 38 (0.03\%) & - & - & - & - & - & - \\ 
   &  7 & - & - & - & - & - & - & - & - & - \\ 
   &  8 & - & - & - & - & - & - & - & - & - \\ 
   &  9 & - & - & - & - & - & - & - & - & - \\ 
   &  10 & 22119 (16.81\%) & 71349 (50.49\%) & 55336 (43.05\%) & 7547 (7.07\%) & 592 (0.64\%) & 2 (0\%) & - & - & - \\ \cline{2-11}
   &  Total & 23751 & 73909 & 56109 & 7557 & 592 & 2 & 0 & 0 & 0 \\ 
   & Never cross & 14996 (63.14\%) & 59809 (80.92\%) & 47403 (84.48\%) & 3593 (47.55\%) & 507 (85.64\%) & - & - & - & - \\ 
  \midrule
  \parbox[t]{10mm}{\multirow{12}{*}{\rotatebox[origin=c]{90}{Low}}}   &  1 & - & - & - & - & - & - & - & - & - \\ 
   &  2 & - & - & - & - & - & - & - & - & - \\ 
   &  3 & - & - & - & - & - & - & - & - & - \\ 
   &  4 & - & - & - & - & - & - & - & - & - \\ 
   &  5 & - & - & - & - & - & - & - & - & - \\ 
   &  6 & - & - & - & - & - & - & - & - & - \\ 
   &  7 & - & - & - & - & - & - & - & - & - \\ 
   &  8 & - & - & - & - & - & - & - & - & - \\ 
   &  9 & - & - & - & - & - & - & - & - & - \\ 
   &  10 & 102989 (78.29\%) & 44529 (31.51\%) & 5541 (4.31\%) & 149 (0.14\%) & 3 (0\%) & - & - & - & - \\
   \cline{2-11}
   &  Total & 102989 & 44529 & 5541 & 149 & 3 & 0 & 0 & 0 & 0 \\ 
   & Never cross & 100361 (97.45\%) & 44111 (99.06\%) & 5533 (99.86\%) & 114 (76.51\%) & 3 (100\%) & - & - & - & - \\ 
   \midrule
   & Total & 131548 & 141304 & 128547 & 106788 & 92200 & 67790 & 45512 & 30978 & 23189 \\ 
   & Never cross & 115474 (87.78\%) & 105898 (74.94\%) & 55293 (43.01\%) & 5977 (5.6\%) & 1508 (1.64\%) & - & - & - & - \\ 
   \hline
\end{tabular}}
\end{table}

\begin{figure}[ht]
    \centering
    \includegraphics[width = 0.7\textwidth]{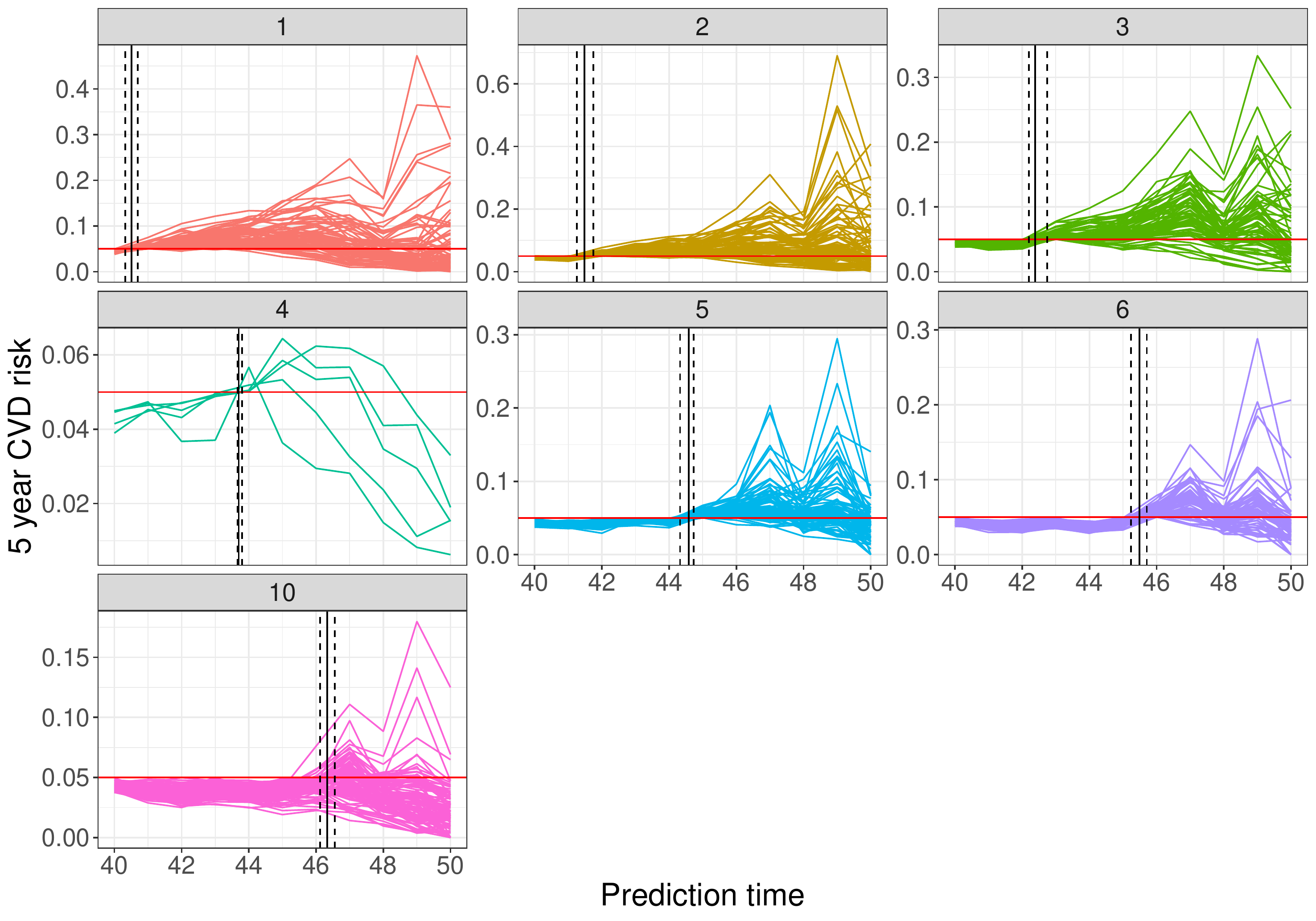}
    \caption{Risk profile representation for women at $L_a=40$. Each panel represents an optimal risk-assessment frequency. Each line is the estimated risk profile for a specific person. The solid black line represents the median expected time of crossing the 5\% threshold (represented via the horizontal red line). The dashed lines represent the first and third quartiles of the $t_{k^*_i,40}$ distribution. }
    \label{fig:risk_profile_women_40_HR}
\end{figure}

In Figure~\ref{fig:risk_profile_women_40_HR} we report a detailed representation of risk profiles, $\hat{r}_i(s + 5;\textbf{x}_i(s), s)$ $s \in \{40,\dots,50\}$, for women at $L_a = 40$, whose 5-year CVD risk at $L_a=40$ is classified as high. It is immediate to notice that the median $\tau_{k^*_{i,40}}$ (black solid lines) increases according to the optimal frequency recommendation. Indeed, people whose 5-year CVD risk is expected to exceed the 5\% threshold later in time are more likely to be recommended a lower frequency risk-assessment strategy. Furthermore, it is interesting to notice that the 5-year CVD risk estimated at the baseline is relevant for the risk-assessment recommendation, but the information given by the risk profile is fundamental for the risk-assessment recommendation. Indeed, we observe that different risk-profile trends (more steep or more flat) can lead to opposite risk-assessment recommendation even for people at high risk of CVD.

\subsection{Descriptive statistics of the landmark cohorts}
In this Section, we report the descriptive characteristics of each landmark cohort, stratified by the 5-year CVD risk classification at the landmark age. Descriptive statistics are reported in table~\ref{tab:women_baseline_characteristics} and~\ref{tab:men_baseline_characteristics} for women and men respectively. We observe a higher risk for people under blood pressure medication, with diagnoses of  depression, diabetes, migraine, renal disease, rheumatoid arthritis, severe mental illness and systemic lupus eritematosus. Higher values of SBP, total cholesterol (TCHOL), BMI are associated with people at higher risk.  The Townsend 20 index and smoking are also associated with higher risk.

\begin{table}[ht!]
\centering
\caption{\\ Descriptive characteristics at each landmark age for the women cohorts.}
\label{tab:women_baseline_characteristics}
\resizebox{\textwidth}{!}{%
\begin{tabular}{ccccccccccc}
  \toprule
variable & Risk class   & 40 & 45 & 50 & 55 & 60 & 65 & 70 & 75 & 80 \\ 
  \midrule
  \parbox[t]{10mm}{\multirow{5}{*}{\rotatebox[origin=c]{90}{AF}}}
 & Very high risk & 0 \% & 0.27 \% & 0.47 \% & 0.67 \% & 5.53 \% & 4.37 \% & 2.8 \% & 3.4 \% & 5.63 \% \\ 
   & High risk & 0.41 \% & 0.14 \% & 0.56 \% & 0.67 \% & 0.65 \% & 0.14 \% & 0 \% & 0 \% &  \\ 
   & Med-high risk & 0.19 \% & 0.26 \% & 0.39 \% & 0.63 \% & 0.1 \% & 0 \% & 0 \% & 0 \% &  \\ 
   & Med-low risk & 0.16 \% & 0.23 \% & 0.36 \% & 0.39 \% & 0 \% & 0 \% &  &  &  \\ 
   & Low risk & 0.12 \% & 0.15 \% & 0.2 \% & 0.29 \% & 0 \% & 0 \% &  &  &  \\ 
 \midrule
  \parbox[t]{10mm}{\multirow{5}{*}{\rotatebox[origin=c]{90}{BP med}}} & Very high risk & 95.82 \% & 91.6 \% & 92.58 \% & 83.88 \% & 80.16 \% & 77.39 \% & 57.62 \% & 52.06 \% & 62.15 \% \\ 
   & High risk & 87.01 \% & 85.62 \% & 81.34 \% & 64.71 \% & 62.07 \% & 52.84 \% & 1.27 \% & 0.48 \% &  \\ 
   & Med-high risk & 78.43 \% & 72.31 \% & 65.12 \% & 53.63 \% & 39.6 \% & 16.35 \% & 0 \% & 0 \% &  \\ 
   & Med-low risk & 55.01 \% & 44.93 \% & 31.13 \% & 22.5 \% & 8.73 \% & 1.63 \% &  &  &  \\ 
   & Low risk & 7.61 \% & 7.47 \% & 3.43 \% & 3.38 \% & 0.2 \% & 0 \% &  &  &  \\
   \midrule
  \parbox[t]{10mm}{\multirow{5}{*}{\rotatebox[origin=c]{90}{Depression}}} & Very high risk & 83.29 \% & 73.19 \% & 82.12 \% & 72.97 \% & 64.78 \% & 43.67 \% & 28.51 \% & 20.35 \% & 19.39 \% \\ 
   & High risk & 75.05 \% & 64.02 \% & 71.53 \% & 64.7 \% & 47.21 \% & 29.05 \% & 4.3 \% & 2.17 \% &  \\ 
   & Med-high risk & 75.19 \% & 63.24 \% & 61.72 \% & 49.15 \% & 32.24 \% & 18.46 \% & 0.63 \% & 0 \% &  \\ 
   & Med-low risk & 66.92 \% & 52.9 \% & 44.86 \% & 28.02 \% & 11.96 \% & 7 \% &  &  &  \\ 
   & Low risk & 26.21 \% & 25.03 \% & 14.42 \% & 7.79 \% & 0.97 \% & 0 \% &  &  &  \\
   \midrule
  \parbox[t]{10mm}{\multirow{5}{*}{\rotatebox[origin=c]{90}{Diabetes}}}& Very high risk & 22.84 \% & 21.18 \% & 11.88 \% & 13.12 \% & 6.22 \% & 4.31 \% & 2.07 \% & 1.96 \% & 2.11 \% \\ 
   & High risk & 15.67 \% & 11.52 \% & 4.91 \% & 3.47 \% & 1.41 \% & 0.37 \% & 0 \% & 0 \% &  \\ 
   & Med-high risk & 6.92 \% & 5.09 \% & 2.28 \% & 1.06 \% & 0.43 \% & 0.09 \% & 0 \% & 0 \% &  \\ 
   & Med-low risk & 3.54 \% & 1.91 \% & 0.76 \% & 0.11 \% & 0.07 \% & 0 \% &  &  &  \\ 
   & Low risk & 0.31 \% & 0.15 \% & 0.06 \% & 0.01 \% & 0 \% & 0 \% &  &  &  \\
   \midrule
  \parbox[t]{10mm}{\multirow{5}{*}{\rotatebox[origin=c]{90}{Migraine}}} & Very high risk & 62.95 \% & 52.37 \% & 41.27 \% & 26.2 \% & 20.06 \% & 15.24 \% & 7.61 \% & 6.7 \% & 5.19 \% \\ 
   & High risk & 52.58 \% & 46.28 \% & 33.12 \% & 21.57 \% & 16.63 \% & 11.44 \% & 8.68 \% & 0.97 \% &  \\ 
   & Med-high risk & 54.57 \% & 41.99 \% & 27.26 \% & 18.33 \% & 12.51 \% & 8.08 \% & 25.83 \% & 0 \% &  \\ 
   & Med-low risk & 42.54 \% & 30.5 \% & 16.82 \% & 12 \% & 6.93 \% & 3.47 \% &  &  &  \\ 
   & Low risk & 9.99 \% & 8 \% & 6.11 \% & 4.97 \% & 2.8 \% & 0 \% &  &  & \\
   \midrule
  \parbox[t]{10mm}{\multirow{5}{*}{\rotatebox[origin=c]{90}{Renal dis.}}}& Very high risk & 3.06 \% & 2.59 \% & 3.04 \% & 3.18 \% & 5.09 \% & 9.32 \% & 6.86 \% & 7.86 \% & 12.18 \% \\ 
   & High risk & 1.86 \% & 1.79 \% & 2.68 \% & 2.76 \% & 3.4 \% & 3.03 \% & 0.97 \% & 0 \% &  \\ 
   & Med-high risk & 1.46 \% & 2.19 \% & 2.22 \% & 2.06 \% & 2.13 \% & 1.15 \% & 0.09 \% & 0 \% &  \\ 
   & Med-low risk & 1.2 \% & 1.24 \% & 1.29 \% & 1.3 \% & 1.18 \% & 0.32 \% &  &  &  \\ 
   & Low risk & 0.36 \% & 0.5 \% & 0.67 \% & 0.93 \% & 1.26 \% & 0 \% &  &  &  \\
   \midrule
  \parbox[t]{10mm}{\multirow{5}{*}{\rotatebox[origin=c]{90}{RA}}} & Very high risk & 2.23 \% & 1.52 \% & 1.78 \% & 2.41 \% & 8.52 \% & 6.64 \% & 3.53 \% & 2.89 \% & 3.03 \% \\ 
   & High risk & 1.03 \% & 1.14 \% & 1.62 \% & 2.08 \% & 3.3 \% & 1.94 \% & 0.03 \% & 0.48 \% &  \\ 
   & Med-high risk & 0.7 \% & 1.52 \% & 1.64 \% & 1.81 \% & 1.51 \% & 0.43 \% & 0 \% & 0 \% &  \\ 
   & Med-low risk & 0.89 \% & 1.01 \% & 1.27 \% & 1.44 \% & 0.2 \% & 0.03 \% &  &  &  \\ 
   & Low risk & 0.66 \% & 0.86 \% & 1.05 \% & 1.42 \% & 0 \% & 0 \% &  &  &  \\
   \midrule
  \parbox[t]{10mm}{\multirow{5}{*}{\rotatebox[origin=c]{90}{SMI}}} & Very high risk & 9.75 \% & 6.43 \% & 9.82 \% & 8.07 \% & 5.19 \% & 4.52 \% & 1.99 \% & 1.61 \% & 1.73 \% \\ 
   & High risk & 7.01 \% & 4.94 \% & 5.92 \% & 3.89 \% & 2.54 \% & 0.96 \% & 0.02 \% & 0 \% &  \\ 
   & Med-high risk & 8.38 \% & 4.25 \% & 3.93 \% & 2.43 \% & 1.31 \% & 0.22 \% & 0 \% & 0 \% &  \\ 
   & Med-low risk & 5.05 \% & 2.9 \% & 1.72 \% & 0.96 \% & 0.32 \% & 0.01 \% &  &  &  \\ 
   & Low risk & 0.78 \% & 0.81 \% & 0.23 \% & 0.2 \% & 0.03 \% & 0 \% &  &  &  \\
   \midrule
  \parbox[t]{10mm}{\multirow{5}{*}{\rotatebox[origin=c]{90}{SLE}}} & Very high risk & 1.11 \% & 0.89 \% & 0.45 \% & 0.41 \% & 0.42 \% & 0.44 \% & 0.3 \% & 0.32 \% & 0.23 \% \\ 
   & High risk & 0 \% & 0.36 \% & 0.35 \% & 0.46 \% & 0.38 \% & 0.32 \% & 0.2 \% & 0.24 \% &  \\ 
   & Med-high risk & 0.57 \% & 0.44 \% & 0.41 \% & 0.37 \% & 0.29 \% & 0.21 \% & 0.18 \% & 0 \% &  \\ 
   & Med-low risk & 0.38 \% & 0.37 \% & 0.34 \% & 0.27 \% & 0.23 \% & 0.27 \% &  &  &  \\ 
   & Low risk & 0.19 \% & 0.22 \% & 0.23 \% & 0.25 \% & 0.4 \% & 0 \% &  &  &  \\
   \midrule
  \parbox[t]{10mm}{\multirow{5}{*}{\rotatebox[origin=c]{90}{BMI}}}  & Very high risk & 1.22 (1.13) & 0.89 (1.04) & 0.33 (0.86) & 0.08 (0.77) & 0.01 (0.75) & -0.14 (0.63) & -0.27 (0.53) & -0.37 (0.49) & -0.46 (0.44) \\ 
   & High risk & 0.88 (1.03) & 0.58 (0.94) & 0.16 (0.78) & -0.04 (0.68) & -0.11 (0.6) & -0.24 (0.52) & -0.45 (0.4) & -0.78 (0.46) &  \\ 
   & Med-high risk & 0.54 (0.94) & 0.36 (0.86) & 0.04 (0.72) & -0.1 (0.62) & -0.21 (0.51) & -0.32 (0.44) & -0.74 (0.4) & -1.18 (0.52) &  \\ 
   & Med-low risk & 0.23 (0.83) & 0.05 (0.71) & -0.14 (0.59) & -0.21 (0.52) & -0.34 (0.41) & -0.48 (0.42) &  &  &  \\ 
   & Low risk & -0.33 (0.48) & -0.32 (0.47) & -0.34 (0.45) & -0.39 (0.44) & -0.66 (0.39) & -0.97 (0.3) &  &  &  \\
   \midrule
  \parbox[t]{10mm}{\multirow{5}{*}{\rotatebox[origin=c]{90}{HDL}}} & Very high risk & -1.08 (0.44) & -0.7 (0.57) & -0.46 (0.57) & -0.32 (0.57) & -0.03 (0.61) & 0.01 (0.59) & 0.2 (0.59) & 0.29 (0.59) & 0.3 (0.58) \\ 
   & High risk & -0.86 (0.48) & -0.57 (0.54) & -0.28 (0.55) & -0.14 (0.52) & 0.09 (0.56) & 0.19 (0.54) & 0.44 (0.56) & 1.77 (0.68) &  \\ 
   & Med-high risk & -0.68 (0.47) & -0.43 (0.54) & -0.16 (0.53) & -0.01 (0.49) & 0.19 (0.53) & 0.3 (0.52) & 1.12 (0.82) & 2.52 (0.28) &  \\ 
   & Med-low risk & -0.48 (0.42) & -0.26 (0.46) & -0.02 (0.47) & 0.17 (0.45) & 0.33 (0.51) & 0.57 (0.65) &  &  &  \\ 
   & Low risk & -0.12 (0.33) & 0.01 (0.4) & 0.2 (0.46) & 0.46 (0.57) & 0.72 (0.71) & 1.91 (0.76) &  &  &  \\
   \midrule
  \parbox[t]{10mm}{\multirow{5}{*}{\rotatebox[origin=c]{90}{SBP}}} & Very high risk & 0.19 (0.67) & 0.23 (0.64) & 0.04 (0.6) & 0.05 (0.55) & 0.15 (0.51) & 0.18 (0.46) & 0.2 (0.4) & 0.26 (0.39) & 0.33 (0.42) \\ 
   & High risk & 0 (0.57) & 0.01 (0.58) & -0.1 (0.54) & -0.07 (0.5) & 0.01 (0.43) & 0.06 (0.37) & 0.02 (0.3) & -0.02 (0.36) &  \\ 
   & Med-high risk & -0.24 (0.55) & -0.17 (0.53) & -0.22 (0.51) & -0.15 (0.46) & -0.07 (0.37) & -0.02 (0.33) & -0.35 (0.32) & -0.32 (0.24) &  \\ 
   & Med-low risk & -0.47 (0.47) & -0.4 (0.45) & -0.37 (0.42) & -0.26 (0.38) & -0.2 (0.34) & -0.2 (0.34) &  &  &  \\ 
   & Low risk & -0.85 (0.33) & -0.69 (0.34) & -0.54 (0.35) & -0.44 (0.36) & -0.58 (0.33) & -0.78 (0.23) &  &  &  \\
   \midrule
  \parbox[t]{10mm}{\multirow{5}{*}{\rotatebox[origin=c]{90}{Smoke}}} & Very high risk & 0.71 (0.24) & 0.71 (0.24) & 0.68 (0.26) & 0.72 (0.24) & 0.51 (0.3) & 0.4 (0.29) & 0.26 (0.23) & 0.2 (0.19) & 0.18 (0.19) \\ 
   & High risk & 0.66 (0.25) & 0.68 (0.24) & 0.6 (0.26) & 0.61 (0.25) & 0.42 (0.26) & 0.29 (0.22) & 0.17 (0.1) & 0.08 (0.08) &  \\ 
   & Med-high risk & 0.64 (0.24) & 0.65 (0.24) & 0.56 (0.26) & 0.51 (0.25) & 0.35 (0.21) & 0.24 (0.14) & 0.07 (0.08) & 0.07 (0.08) &  \\ 
   & Med-low risk & 0.61 (0.24) & 0.6 (0.24) & 0.5 (0.24) & 0.38 (0.18) & 0.28 (0.15) & 0.18 (0.12) &  &  &  \\ 
   & Low risk & 0.46 (0.2) & 0.43 (0.2) & 0.38 (0.18) & 0.27 (0.16) & 0.14 (0.13) & 0.04 (0.06) &  &  &  \\
   \midrule
  \parbox[t]{10mm}{\multirow{5}{*}{\rotatebox[origin=c]{90}{TCHOL}}} & Very high risk & 0.07 (0.56) & 0.34 (0.67) & 0.3 (0.6) & 0.42 (0.6) & 0.44 (0.53) & 0.43 (0.51) & 0.39 (0.46) & 0.36 (0.46) & 0.3 (0.48) \\ 
   & High risk & 0.01 (0.56) & 0.15 (0.57) & 0.22 (0.51) & 0.38 (0.5) & 0.43 (0.44) & 0.44 (0.43) & 0.44 (0.39) & 0.31 (0.55) &  \\ 
   & Med-high risk & -0.1 (0.48) & 0.09 (0.5) & 0.16 (0.44) & 0.33 (0.42) & 0.42 (0.4) & 0.43 (0.38) & 0.43 (0.48) & 0.12 (0.38) &  \\ 
   & Med-low risk & -0.22 (0.4) & -0.05 (0.38) & 0.09 (0.38) & 0.28 (0.36) & 0.39 (0.35) & 0.39 (0.39) &  &  &  \\ 
   & Low risk & -0.38 (0.24) & -0.21 (0.27) & 0.01 (0.31) & 0.21 (0.35) & 0.24 (0.42) & 0.04 (0.57) &  &  &  \\
   \midrule
  \parbox[t]{10mm}{\multirow{5}{*}{\rotatebox[origin=c]{90}{Townsend20}}} & Very high risk & 15.64 (3.66) & 15.01 (4.15) & 14.75 (4.16) & 13.64 (4.73) & 12.97 (4.8) & 11.68 (5.05) & 9.44 (5.19) & 8.69 (5.19) & 9.01 (5.2) \\ 
   & High risk & 15.45 (3.74) & 14.35 (4.46) & 13.28 (4.82) & 12.71 (4.96) & 11.32 (5.1) & 9.22 (5.17) & 5.58 (3.79) & 4.28 (2.97) &  \\ 
   & Med-high risk & 14.97 (4.2) & 13.91 (4.69) & 12.4 (5.15) & 11.29 (5.31) & 9.59 (5.13) & 7.61 (4.41) & 3.65 (2.64) & 2 (1.28) &  \\ 
   & Med-low risk & 14.09 (4.6) & 12.93 (5.09) & 11.05 (5.4) & 9.18 (5.06) & 6.32 (4.12) & 4.24 (3.1) &  &  &  \\ 
   & Low risk & 9.02 (5.52) & 8.31 (5.31) & 7.04 (4.75) & 5.38 (3.97) & 3.1 (2.24) & 2.24 (1.48) &  &  &  \\ 
   \bottomrule
\end{tabular}}
\end{table}

\begin{table}[ht!]
\centering
\caption{\\ \textit{Descriptive characteristics at each landmark age for the men cohorts.}}
\label{tab:men_baseline_characteristics}
\resizebox{\textwidth}{!}{%
\begin{tabular}{ccccccccccc}
  \toprule
variable & Risk class   & 40 & 45 & 50 & 55 & 60 & 65 & 70 & 75 & 80 \\ 
  \midrule
  \parbox[t]{10mm}{\multirow{5}{*}{\rotatebox[origin=c]{90}{AF}}} & Very high risk & 1.18 \% & 0.78 \% & 1.01 \% & 1.4 \% & 2.42 \% & 2.35 \% & 3.42 \% & 5.45 \% & 8.27 \% \\ 
   & High risk & 0.61 \% & 1.09 \% & 0.96 \% & 0.81 \% & 0.01 \% & 0 \% & 0 \% &  &  \\ 
   & Med-high risk & 1.04 \% & 0.73 \% & 0.54 \% & 0.63 \% & 0.01 \% & 0 \% &  &  &  \\ 
   & Med-low risk & 0.45 \% & 0.33 \% & 0.41 \% & 0.69 \% & 0 \% & 0 \% &  &  &  \\ 
   & Low risk & 0.2 \% & 0.29 \% & 0.32 \% & 0.67 \% & 0 \% &  &  &  &  \\ 
  \midrule
  \parbox[t]{10mm}{\multirow{5}{*}{\rotatebox[origin=c]{90}{BP med}}} & Very high risk & 84.23 \% & 80.23 \% & 59.76 \% & 50.48 \% & 36.44 \% & 31.47 \% & 36.19 \% & 44.84 \% & 55.08 \% \\ 
   & High risk & 78.51 \% & 56.12 \% & 30.35 \% & 11.88 \% & 5.5 \% & 2.37 \% & 0 \% &  &  \\ 
   & Med-high risk & 61.79 \% & 29.84 \% & 9.59 \% & 3.13 \% & 3.85 \% & 1.43 \% &  &  &  \\ 
   & Med-low risk & 18.26 \% & 4.46 \% & 2.22 \% & 2.17 \% & 2.53 \% & 0 \% &  &  &  \\ 
   & Low risk & 1.13 \% & 0.84 \% & 1.08 \% & 0 \% & 0 \% &  &  &  &  \\ 
  \midrule
  \parbox[t]{10mm}{\multirow{5}{*}{\rotatebox[origin=c]{90}{Depression}}} & Very high risk & 55.98 \% & 44.94 \% & 39.69 \% & 31.7 \% & 21.12 \% & 14.4 \% & 11.64 \% & 10.35 \% & 9.81 \% \\ 
   & High risk & 56.29 \% & 39.13 \% & 27 \% & 14.88 \% & 4.52 \% & 2.94 \% & 3.85 \% &  &  \\ 
   & Med-high risk & 47.88 \% & 33.09 \% & 17.43 \% & 4.83 \% & 3.08 \% & 1.67 \% &  &  &  \\ 
   & Med-low risk & 42.5 \% & 15.17 \% & 6.74 \% & 3.71 \% & 1.86 \% & 0 \% &  &  &  \\ 
   & Low risk & 7.2 \% & 5.39 \% & 3.83 \% & 2.68 \% & 0 \% &  &  &  &  \\ 
  \midrule
  \parbox[t]{10mm}{\multirow{5}{*}{\rotatebox[origin=c]{90}{Diabetes}}} & Very high risk & 25.23 \% & 4.34 \% & 6.81 \% & 4.17 \% & 2.92 \% & 2.45 \% & 2.94 \% & 3.7 \% & 3.82 \% \\ 
   & High risk & 8.7 \% & 3.44 \% & 1.68 \% & 0.78 \% & 0.16 \% & 0.28 \% & 0 \% &  &  \\ 
   & Med-high risk & 5.64 \% & 1.79 \% & 0.61 \% & 0.29 \% & 0.09 \% & 1.19 \% &  &  &  \\ 
   & Med-low risk & 1.1 \% & 0.59 \% & 0.16 \% & 0.17 \% & 0.51 \% & 0 \% &  &  &  \\ 
   & Low risk & 0.08 \% & 0.46 \% & 0.18 \% & 1.34 \% & 0 \% &  &  &  &  \\ 
  \midrule
  \parbox[t]{10mm}{\multirow{5}{*}{\rotatebox[origin=c]{90}{Migraine}}} & Very high risk & 8.67 \% & 17.41 \% & 10.62 \% & 7.6 \% & 4.75 \% & 3.45 \% & 3.16 \% & 2.63 \% & 2.27 \% \\ 
   & High risk & 10.95 \% & 14.85 \% & 6.32 \% & 4.15 \% & 2.22 \% & 3.41 \% & 0 \% &  &  \\ 
   & Med-high risk & 9.15 \% & 10.21 \% & 4.54 \% & 1.57 \% & 1.93 \% & 2.38 \% &  &  &  \\ 
   & Med-low risk & 6.18 \% & 3.66 \% & 1.86 \% & 1.01 \% & 2.03 \% & 0 \% &  &  &  \\ 
   & Low risk & 3.84 \% & 1.06 \% & 1.3 \% & 0 \% & 0 \% &  &  &  &  \\ 
  \midrule
  \parbox[t]{10mm}{\multirow{5}{*}{\rotatebox[origin=c]{90}{Renal dis}}} & Very high risk & 3.42 \% & 1.91 \% & 1.89 \% & 1.76 \% & 2.19 \% & 2.26 \% & 3.66 \% & 6.29 \% & 10.26 \% \\ 
   & High risk & 1.74 \% & 1.95 \% & 1.38 \% & 0.86 \% & 0.27 \% & 2.72 \% & 0 \% &  &  \\ 
   & Med-high risk & 1.47 \% & 1.04 \% & 0.64 \% & 0.62 \% & 0.2 \% & 3.1 \% &  &  &  \\ 
   & Med-low risk & 0.7 \% & 0.42 \% & 0.5 \% & 0.98 \% & 0.34 \% & 50 \% &  &  &  \\ 
   & Low risk & 0.25 \% & 0.38 \% & 0.83 \% & 0 \% & 0 \% &  &  &  &  \\ 
  \midrule
  \parbox[t]{10mm}{\multirow{5}{*}{\rotatebox[origin=c]{90}{RA}}} & Very high risk & 0.66 \% & 0.69 \% & 0.77 \% & 0.79 \% & 1.3 \% & 1.09 \% & 1.24 \% & 1.38 \% & 1.44 \% \\ 
   & High risk & 0.51 \% & 0.55 \% & 0.66 \% & 0.59 \% & 0.03 \% & 0.09 \% & 0 \% &  &  \\ 
   & Med-high risk & 0.55 \% & 0.55 \% & 0.5 \% & 0.52 \% & 0.04 \% & 0 \% &  &  &  \\ 
   & Med-low risk & 0.39 \% & 0.41 \% & 0.47 \% & 0.91 \% & 0 \% & 0 \% &  &  &  \\ 
   & Low risk & 0.31 \% & 0.42 \% & 0.7 \% & 0.67 \% & 0 \% &  &  &  &  \\ 
  \midrule
  \parbox[t]{10mm}{\multirow{5}{*}{\rotatebox[origin=c]{90}{SMI}}} & Very high risk & 6.44 \% & 5.96 \% & 3.73 \% & 2.3 \% & 1.92 \% & 1.22 \% & 1.2 \% & 1.14 \% & 1.11 \% \\ 
   & High risk & 4.09 \% & 5.44 \% & 2.17 \% & 1.15 \% & 0.14 \% & 0.14 \% & 0 \% &  &  \\ 
   & Med-high risk & 3.78 \% & 3.4 \% & 1.33 \% & 0.6 \% & 0.08 \% & 0.24 \% &  &  &  \\ 
   & Med-low risk & 2.74 \% & 1.16 \% & 0.63 \% & 0.81 \% & 0.17 \% & 0 \% &  &  &  \\ 
   & Low risk & 1.17 \% & 0.49 \% & 0.76 \% & 0.67 \% & 0 \% &  &  &  &  \\ 
  \midrule
  \parbox[t]{10mm}{\multirow{5}{*}{\rotatebox[origin=c]{90}{SLE}}} & Very high risk & 0 \% & 0.1 \% & 0.09 \% & 0.09 \% & 0.07 \% & 0.08 \% & 0.09 \% & 0.09 \% & 0.09 \% \\ 
   & High risk & 0.2 \% & 0.02 \% & 0.1 \% & 0.05 \% & 0.04 \% & 0.14 \% & 0 \% &  &  \\ 
   & Med-high risk & 0 \% & 0.09 \% & 0.05 \% & 0.06 \% & 0.07 \% & 0.24 \% &  &  &  \\ 
   & Med-low risk & 0.05 \% & 0.05 \% & 0.04 \% & 0.08 \% & 0 \% & 0 \% &  &  &  \\ 
   & Low risk & 0.04 \% & 0.02 \% & 0.07 \% & 0 \% & 0 \% &  &  &  &  \\ 
  \midrule
  \parbox[t]{10mm}{\multirow{5}{*}{\rotatebox[origin=c]{90}{BMI}}} & Very high risk & 0.78 (1.09) & 0.45 (0.94) & 0.09 (0.73) & -0.1 (0.62) & -0.22 (0.53) & -0.31 (0.49) & -0.4 (0.49) & -0.46 (0.47) & -0.56 (0.43) \\ 
   & High risk & 0.38 (0.91) & 0.16 (0.7) & -0.14 (0.56) & -0.22 (0.44) & -0.32 (0.37) & -0.62 (0.44) & -1.15 (0.5) &  &  \\ 
   & Med-high risk & 0.21 (0.77) & -0.05 (0.56) & -0.18 (0.44) & -0.26 (0.39) & -0.5 (0.45) & -0.89 (0.43) &  &  &  \\ 
   & Med-low risk & -0.11 (0.53) & -0.19 (0.37) & -0.25 (0.37) & -0.49 (0.46) & -0.82 (0.43) & -1.02 (0.78) &  &  &  \\ 
   & Low risk & -0.28 (0.32) & -0.32 (0.38) & -0.52 (0.45) & -0.94 (0.43) & -1.47 (0.18) &  &  &  &  \\ 
  \midrule
  \parbox[t]{10mm}{\multirow{5}{*}{\rotatebox[origin=c]{90}{HDL}}} & Very high risk & -0.7 (0.51) & -0.41 (0.52) & -0.23 (0.51) & -0.07 (0.5) & 0.08 (0.5) & 0.2 (0.56) & 0.28 (0.63) & 0.28 (0.64) & 0.3 (0.62) \\ 
   & High risk & -0.51 (0.46) & -0.28 (0.48) & -0.07 (0.43) & 0.08 (0.41) & 0.25 (0.46) & 0.79 (0.82) & 1.91 (1.3) &  &  \\ 
   & Med-high risk & -0.39 (0.44) & -0.18 (0.41) & -0.01 (0.37) & 0.17 (0.43) & 0.63 (0.76) & 1.78 (1.09) &  &  &  \\ 
   & Med-low risk & -0.22 (0.32) & -0.07 (0.3) & 0.1 (0.41) & 0.68 (0.81) & 1.73 (1.01) & 2.52 (0.98) &  &  &  \\ 
   & Low risk & -0.07 (0.31) & 0.11 (0.48) & 0.66 (0.83) & 2.36 (0.93) & 4.04 (0.08) &  &  &  &  \\ 
  \midrule
  \parbox[t]{10mm}{\multirow{5}{*}{\rotatebox[origin=c]{90}{SBP}}} & Very high risk & 0.17 (0.65) & 0.13 (0.59) & -0.08 (0.54) & -0.08 (0.5) & -0.04 (0.44) & -0.03 (0.41) & -0.01 (0.44) & 0.03 (0.45) & 0.05 (0.46) \\ 
   & High risk & -0.11 (0.54) & -0.15 (0.47) & -0.28 (0.41) & -0.21 (0.34) & -0.17 (0.32) & -0.41 (0.37) & -0.66 (0.42) &  &  \\ 
   & Med-high risk & -0.23 (0.5) & -0.33 (0.38) & -0.33 (0.33) & -0.29 (0.32) & -0.39 (0.38) & -0.59 (0.42) &  &  &  \\ 
   & Med-low risk & -0.46 (0.34) & -0.44 (0.27) & -0.41 (0.31) & -0.49 (0.37) & -0.57 (0.42) & -0.56 (0.37) &  &  &  \\ 
   & Low risk & -0.58 (0.25) & -0.58 (0.29) & -0.62 (0.35) & -0.66 (0.37) & -0.58 (0.59) &  &  &  &  \\ 
  \midrule
  \parbox[t]{10mm}{\multirow{5}{*}{\rotatebox[origin=c]{90}{Smoke}}} & Very high risk & 0.7 (0.22) & 0.68 (0.23) & 0.65 (0.25) & 0.54 (0.28) & 0.4 (0.26) & 0.29 (0.25) & 0.22 (0.23) & 0.17 (0.22) & 0.14 (0.21) \\ 
   & High risk & 0.68 (0.22) & 0.64 (0.23) & 0.61 (0.25) & 0.42 (0.2) & 0.28 (0.15) & 0.15 (0.15) & 0.06 (0.08) &  &  \\ 
   & Med-high risk & 0.65 (0.21) & 0.65 (0.23) & 0.5 (0.19) & 0.34 (0.17) & 0.17 (0.16) & 0.11 (0.12) &  &  &  \\ 
   & Med-low risk & 0.66 (0.19) & 0.54 (0.17) & 0.39 (0.17) & 0.2 (0.19) & 0.14 (0.15) & 0.04 (0.02) &  &  &  \\ 
   & Low risk & 0.52 (0.18) & 0.38 (0.2) & 0.18 (0.18) & 0.15 (0.15) & 0.13 (0.15) &  &  &  &  \\ 
  \midrule
  \parbox[t]{10mm}{\multirow{5}{*}{\rotatebox[origin=c]{90}{TCHOL}}} & Very high risk & 0.68 (0.75) & 0.59 (0.59) & 0.52 (0.52) & 0.45 (0.48) & 0.39 (0.42) & 0.3 (0.41) & 0.19 (0.43) & 0.08 (0.44) & 0.02 (0.45) \\ 
   & High risk & 0.5 (0.56) & 0.5 (0.48) & 0.44 (0.42) & 0.41 (0.36) & 0.36 (0.32) & 0.16 (0.46) & -0.45 (0.64) &  &  \\ 
   & Med-high risk & 0.47 (0.51) & 0.43 (0.4) & 0.42 (0.33) & 0.37 (0.33) & 0.23 (0.44) & 0.02 (0.57) &  &  &  \\ 
   & Med-low risk & 0.33 (0.33) & 0.37 (0.28) & 0.36 (0.32) & 0.21 (0.46) & 0.03 (0.53) & -1.1 (0.51) &  &  &  \\ 
   & Low risk & 0.24 (0.24) & 0.27 (0.33) & 0.17 (0.45) & 0.03 (0.61) & 0.09 (0.25) &  &  &  &  \\ 
  \midrule
  \parbox[t]{10mm}{\multirow{5}{*}{\rotatebox[origin=c]{90}{Townsend 20}}} & Very high risk & 13.97 (4.83) & 12.85 (5.17) & 12.81 (5.16) & 11.26 (5.35) & 10.34 (5.25) & 8.7 (5.22) & 8.39 (5.17) & 8.53 (5.2) & 8.73 (5.21) \\ 
   & High risk & 13.63 (4.75) & 12.2 (5.38) & 11.61 (5.34) & 10.22 (5.31) & 5.96 (4.02) & 4.8 (3.7) & 3.23 (2.41) &  &  \\ 
   & Med-high risk & 12.9 (5.18) & 12.2 (5.36) & 10.92 (5.25) & 6.42 (4.41) & 4.98 (3.86) & 4.4 (3.47) &  &  &  \\ 
   & Med-low risk & 12.9 (5.09) & 10.29 (5.16) & 6.48 (4.44) & 5.35 (4.11) & 4.55 (3.61) & 6.5 (0.71) &  &  &  \\ 
   & Low risk & 8.65 (5.41) & 5.98 (4.81) & 4.96 (4) & 4.72 (4.23) & 5 (1.73) &  &  &  &  \\ 
    \bottomrule
\end{tabular}}
\end{table}




\section{Validation: C-index and Brier score}
\label{sec:gof}

\subsection{Validation}
\label{ss:results_valid}

\begin{figure}[ht]
    \centering
    \includegraphics[width = 0.7\textwidth]{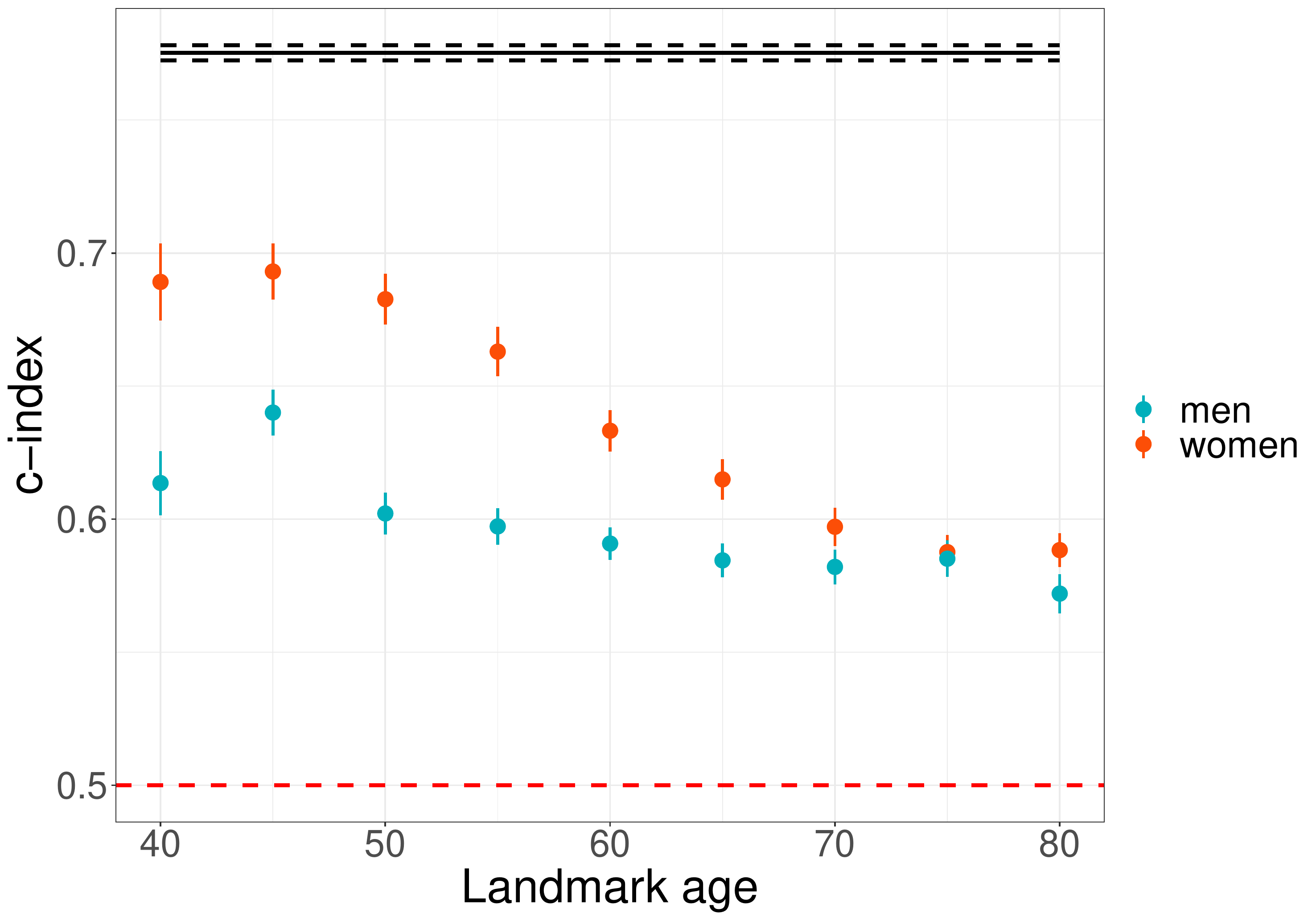}
    \caption{Estimated c-indices for the second landmark model for men (blue dots) and women (red dots) for different values of starting time $t$. Each point represents a c-index computed for a specific $s = L_a \in \{40,45,..,80\}$ and $w=10$, since we are interested in the discrimination accuracy of the 10-year CVD risk. Points at $L_a=40$, represents the c-indices estimated with $s=40$ and $w=10$. The solid black lines represent the overall c-index across landmark ages and gender (dashed lines represent $95\%$ confidence interval). The dashed red line at 0.5 represents the minimum sensible value of the c-index.}
    \label{fig:cindex_10CVD}
\end{figure}
\begin{figure}[ht]
    \centering
    \includegraphics[width = 0.7\textwidth]{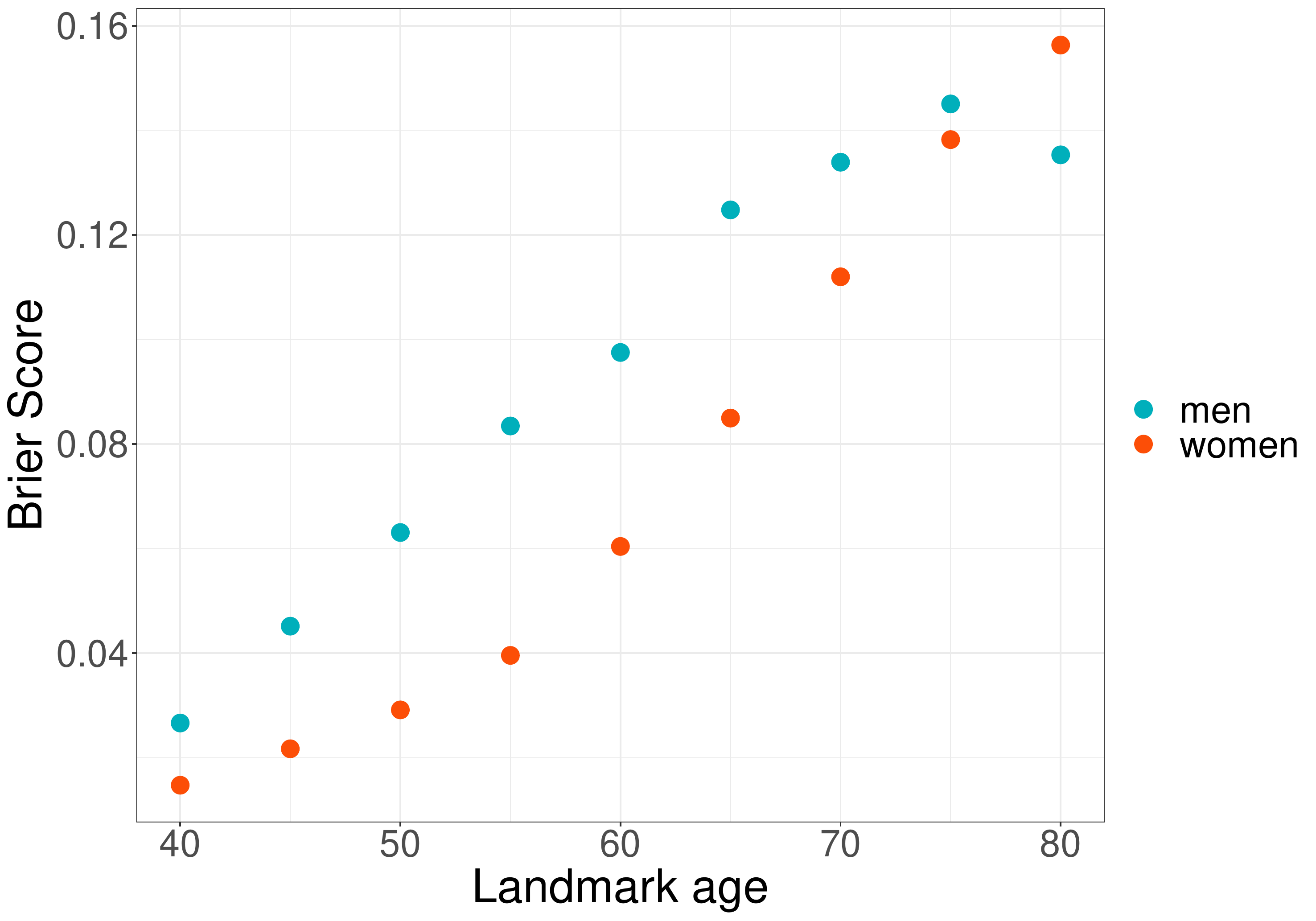}
    \caption{Estimates of Brier Score, $BS_s(w)$ where $s =L_a \in \{40,45,..,80\}$ and $w = 10$. Each $BS_{L_a}(10)$ is represented through a colored dot (blue dots for men and red dots for women).}
    \label{fig:bs_10CVD}
\end{figure}

We validate the 2-stage landmark model for estimating the probability of not being diagnosed with CVD before statins initiation, described in Section 3.2 of the main manuscript. The estimated c-indices are represented via dots in Figure~\ref{fig:cindex_10CVD}, while the Brier scores are represented in Figure~\ref{fig:bs_10CVD}.

In Figure~\ref{fig:cindex_10CVD}, we found good overall discrimination (overall c-index equal to 0.77, represented via a solid black line).

\begin{figure}[ht!]
    \centering
    \includegraphics[width = 0.9\textwidth]{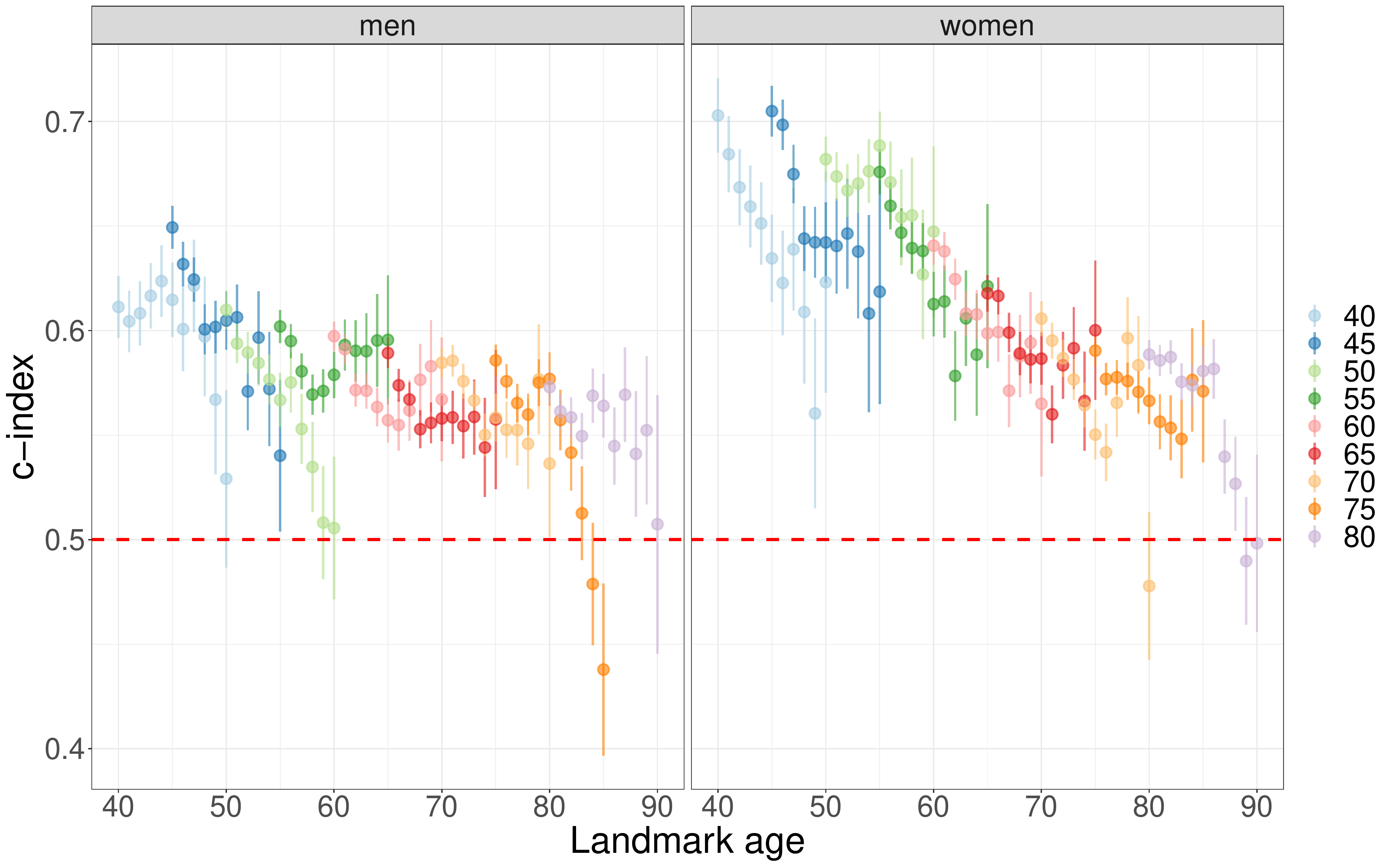}
    \caption{Estimated c-indices for the first landmark model for women (left panel) and men (right panels) for different values of starting time $s$. Each point represents a c-index computed for a specific $s \in \mathcal{P}_{L_a}$ and $w=5$, since we are interested in the discrimination accuracy of the 5-year CVD risk. We associate a specific color to each landmark set. All points in light blue are associated with $L_a = 40$, and from the first point from the left we have $s \in \{40,41,42,..,50\}$.The dashed red line at 0.5 represents the minimum sensible value of the c-index. Values lower than 0.5 are recorded at older ages, for the latest time-windows (i.e., 83-88, 84-89, 85-90 in orange for men, 88-93, 89-94, 90-95 in violet for men).}
    \label{fig:cindex_5CVD}
\end{figure}

\begin{figure}[ht!]
    \centering
    \includegraphics[width = 0.9\textwidth]{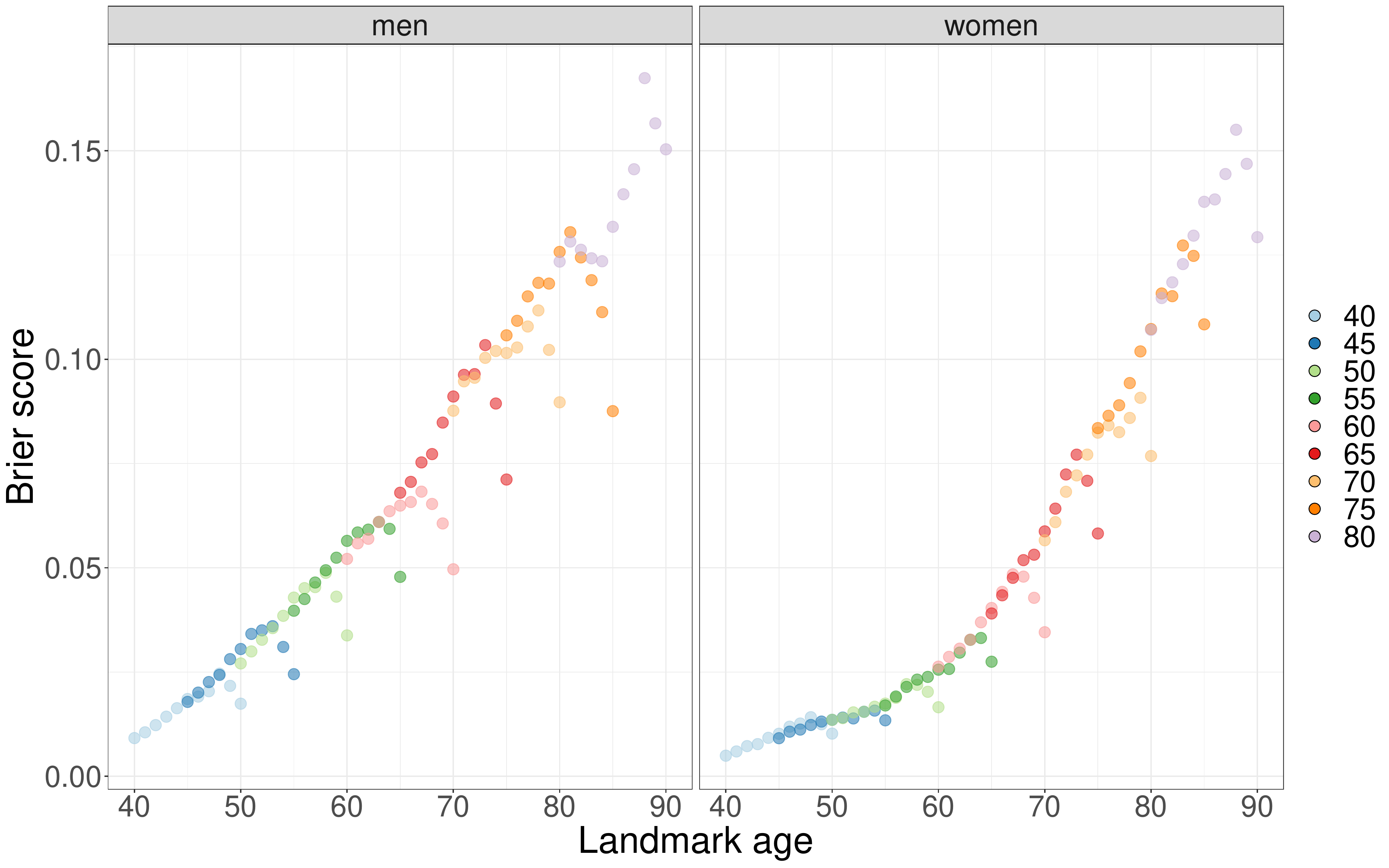}
    \caption{Estimates of Brier Score, $BS_s(w)$ where $s \in \mathcal{P}_{L_a}$ and $w = 5$. Each $BS_s(5)$ is represented through a colored dot (each landmark age is associated to a specific color). BS associated to women are reported in the left panel, while the BS associated to men are reported in the right panel. }
    \label{fig:bs_5CVD}
\end{figure}

Secondly, we validate the extended 2-stage landmarking approach described in Section 3.3 of the main manuscript and we report the estimated c-indices in Figure~\ref{fig:cindex_5CVD} and the estimated Brier Scores in Figure~\ref{fig:bs_5CVD}.  

Note that the extended 2-stage landmarking approach for estimating $t_{i,L_a}^*$ in general has good discriminatory power  and good prediction accuracy (the lower the Brier Score, the higher the predictive accuracy of the model). The model performs better for women than men (at each time $s$ the c-index for women is higher than the c-index for men, and the inverse for the Brier score). Furthermore, the model performance tends to decrease for higher landmark ages. Low c-indices with high standard deviations are found for older landmark ages ($75$, $80$), for later prediction times $s \in \{83,84,85\}$ and $s \in \{88,89,90\}$ respectively. This could be due to the fact that the mean follow-up time is lower at higher landmark ages and very few people are observed after 83 years at landmark age 75 and after 88 years at landmark age 80.

 We observe similar trends between Figure~\ref{fig:cindex_10CVD}, Figure~\ref{fig:bs_10CVD} and  Figure~\ref{fig:cindex_5CVD},  Figure~\ref{fig:bs_5CVD}: lower discrimination and prediction accuracy for the men landmark cohorts and a decline in model performance as the landmark age increases.
 

\section{Sensitivity analysis: exploring the effect of NB parameters}
\label{s:sensitivity}
\begin{figure}
        \centering
        \includegraphics[width = \textwidth]{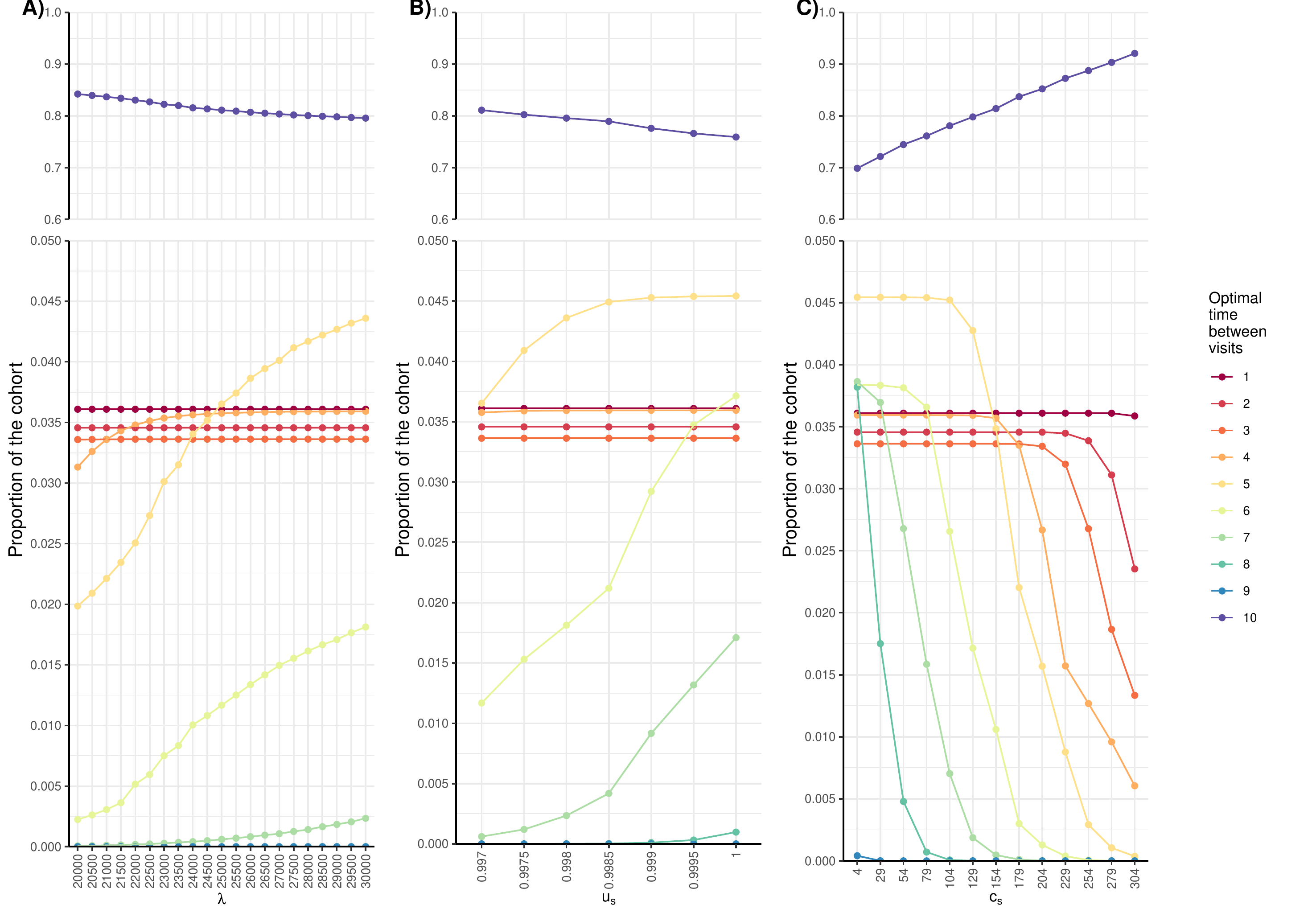}
        \caption{\label{fig:sensitivity}Sensitivity analysis of $\lambda$ (panel A), $c_s$ (panel B), $u_s$ (panel C). In the y-axis of all panels we represent the proportion of the cohort.
        In panel A, the value of $\lambda$ is reported in the x-axis and it ranges in $[20,000; 30,000]$ \pounds/year. $u_{s} = 0.997$, $c_s = 150$ \pounds/year, and $c_{\nu} = 18.39$ \pounds/visit. 
        In panel B, the value of $u_s$ is reported in the x-axis and it ranges in $[0.997; 1]$. $\lambda = 25,000$ \pounds/year, $c_s = 150$ \pounds/year, and $c_{\nu} = 18.39$ \pounds/visit.
        In panel C, the value of $c_s$ is reported in the x-axis and it ranges in $[4; 320]$ \pounds/year. $\lambda = 25,000$ \pounds/year, $u_s = 0.997$, and $c_{\nu} = 18.39$ \pounds/visit. This figure appears in color in the electronic version of this article.}
\end{figure}

We consider three sensitivity analyses by varying key parameters as follows: 

\begin{itemize}
    \item $\lambda \in [20,000; 30,000]$ \pounds/year, while $u_{s} = 0.997$, $c_s = 150$ \pounds/year, and $c_{\nu} = 18.39$ \pounds/visit. The results are reported in panel A of Figure~\ref{fig:sensitivity}.
    \item $u_{s} \in [0.997; 1]$, while $\lambda = 25,000$ \pounds/year, $c_s = 150$ \pounds/year, and $c_{\nu} = 18.39$ \pounds/visit. The results are reported in panel B of Figure~\ref{fig:sensitivity}.
   \item $c_{s} \in [4; 320]$ \pounds/year, while $\lambda = 25,000$ \pounds/year, $u_s = 0.997$, and $c_{\nu} = 18.39$ \pounds/visit. The results are reported in panel C of Figure~\ref{fig:sensitivity}.
\end{itemize}

In general, we observe that results are robust with respect to the parameter choice.

In panel A of Figure~\ref{fig:sensitivity}, we vary the value of $\lambda$ from 20,000 £/year to 30,000 £/year, as $\lambda$ increases, the 10-year frequency is optimal for fewer people, while intermediate frequency (such as 4-7 years) becomes optimal for a larger proportion of people. Visits every 1, 2, 3 years are optimal for a constant proportion of the cohort. An increasing $\lambda$  can be interpreted as a stronger willingness to pay for increased expectancy of CVD-free life years, so intermediate frequencies tend to be preferred over 10-yearly risk-assessments.

We see an analogous behaviour for $u_s$ (panel B of Figure~\ref{fig:sensitivity}). We vary the utility factor from 0.997 (which implies a decrease of quality of life equal to 0.003) to 1 (which implies that taking statins has no effect at all on the quality of life). We notice that 1 to 4-year risk-assessment strategies are the optima for a constant number of people. We note when the impact of statins on quality of life is low ($u_s$ tends to 1), the 10-year frequency schedule is less preferred, while intermediate frequencies (4-8 years) are preferred. If no burden is associated with taking statins ($u_s =1$), then the optimal strategy is to initiate statins immediately.

In contrast, the higher the price of statins, $c_s$, the more risk-assessment strategies associated with less frequent visits are to be preferred (panel C of Figure~\ref{fig:sensitivity}). This is expected because higher costs imply decreased net benefit of statin usage.

We investigated also $c_{\nu}$ varying between 15 £/visit to 1000 £/visit (results not shown). Despite the broad range explored, the optimal schedule proportions are unchanging across all values of $c_{\nu}$. This result is expected because this term of the NB is not comparable in scale with the terms associated with expected event free life years in Eq. (2) of the manuscript. 

It is also immediate to notice from the range reported in the y-axis of Figure~\ref{fig:sensitivity} that the greatest part of the whole cohort ($>$70\%) is recommended to be assessed every 10 years. This is due to the fact that we are considering the stacked landmark cohorts and the biggest landmark cohorts are those ones collected at $L_a = 40$, $L_a=45$, that are composed of younger and healthier people.

\bibliographystyle{plainnat}
\bibliography{biblio}